\documentclass[journal]{IEEEtran}
\usepackage{cite}
\usepackage{bm}
\usepackage{amsmath}
\usepackage{extarrows}
\usepackage{amssymb}
\usepackage{graphicx}
\usepackage{color}
\usepackage{enumerate}
\usepackage{bookmark} 
\graphicspath{{figure}}
\usepackage{setspace}
\usepackage{setspace}

\newcommand{\mb}{\bm}
\newcommand{\mr}{\mathrm}
\newcommand{\ms}{\mathrm}
\newcommand{\BE}{\begin{equation}}
\newcommand{\EE}{\end{equation}}
\newcommand{\BS}{\begin{subequations}}
\newcommand{\ES}{\end{subequations}}
\renewcommand{\bf}{\bm}

\newtheorem{theorem}{Theorem}
\newtheorem{proposition}{Proposition}
\newtheorem{assumption}{Assumption}
\newtheorem{definition}{Definition}

\newtheorem{lemma}{Lemma}

\ifCLASSINFOpdf
\graphicspath{{figure/}}
\else
\fi
\begin{document}

\title{{On} Capacity Optimality of OAMP: Beyond IID Sensing Matrices and Gaussian Signaling} 

\author{\IEEEauthorblockN{Lei~Liu, \emph{Senior Member, IEEE}, Shansuo~Liang, and~Li~Ping, \emph{Fellow, IEEE}}
\thanks{This article has been presented in part at the 2022 IEEE ISIT, Finland, \cite{Cap_OAMPISIT}.}
\thanks{Lei Liu is with the Zhejiang Provincial Key Laboratory of Information Processing, Communication and Networking, College of Information Science and Electronic Engineering, Zhejiang University, Hangzhou 310007, China (e-mail: lei\_liu@zju.edu.cn).}
\thanks{Shansuo~Liang is with the Theory lab, Central Research Institute, 2012 Labs, Huawei Technologies Co., Ltd., Hong Kong, China (e-mail: liang.shansuo@huawei.com).}
\thanks{Li~Ping is with the Department of Electronic Engineering, City University of Hong Kong, Hong Kong, SAR, China (e-mail: eeliping@cityu.edu.hk).}
}

\maketitle 

\begin{abstract}
This paper investigates a large unitarily invariant system (LUIS) involving a unitarily invariant sensing matrix, an arbitrarily fixed signal distribution, and forward error control (FEC) coding. A universal Gram-Schmidt orthogonalization is considered for constructing orthogonal approximate message passing (OAMP), enabling its applicability to a wide range of prototypes without the constraint of differentiability. We develop two single-input-single-output variational transfer functions for OAMP with Lipschitz continuous local estimators, facilitating an analysis of achievable rates. Furthermore, when the state evolution of OAMP has a unique fixed point, we reveal that OAMP can achieve the constrained capacity predicted by the replica method of LUIS based on matched FEC coding, regardless of the signal distribution. The replica method is rigorously validated for LUIS with Gaussian signaling and certain sub-classes of LUIS with arbitrary signal distributions. Several area properties are established based on the variational transfer functions of OAMP.  Meanwhile, we present a replica constrained capacity-achieving coding principle for LUIS. This principle serves as the basis for optimizing irregular low-density parity-check (LDPC) codes specifically tailored for binary signaling in our simulation results. The performance of OAMP with these optimized codes exhibits a remarkable improvement over the unoptimized codes and even surpasses the well-known Turbo-LMMSE algorithm. For quadrature phase-shift keying (QPSK) modulation, we observe bit error rates (BER) performance near the replica constrained capacity across diverse channel conditions.
 \end{abstract}


\section{Introduction}

\subsection{Receiver Optimality in Un-coded \& Coded Linear Systems}
Consider estimating $\bf{x}=\{x_i\}\in\mathbb{C}^{N\times 1}$ from its observation $\bf{y}\in\mathbb{C}^{M\times 1}$ in a linear system
\BE\label{Eqn:linear_system}
    \bf{y}=\bf{Ax}+\bf{n},
\EE
where $\bm{n}\!\sim\!\mathcal{CN}(\mathbf{0},\sigma^2\bm{I}_M)$ is a Gaussian noise vector. For convenience, we refer to $\bf{A}\in\mathbb{C}^{M\times N}$ as a sensing matrix. We assume that $\{x_i\}$ follow an a priori distribution $x_i \sim P_X(x_i),\forall i$.
Furthermore, we assume that $\bf{x}$ is encoded by a forward error control (FEC) code that includes un-coded $\bf{x}$ as a special trivial case. 

A wide range of communication applications can be represented by \eqref{Eqn:linear_system}, including a well-known example known as the multiple-input multiple-output (MIMO) system, where $\bf{A}$ represents the channel coefficient matrix \cite{Biglieri2007,David2005} and $P_X(x_i)$ is determined by the signaling (i.e., modulation) method. While continuous Gaussian signaling, where the variables ${x_i}$ are independently and identically distributed (IID) as Gaussian, is commonly assumed in information theoretical studies, practical implementations require discrete signaling. One such example is quadrature phase-shift keying (QPSK)  uniformly distributed on $\big\{ \! \pm\! \tfrac{1}{\sqrt{2}}\pm \!\tfrac{1}{\sqrt{2}}\!\cdot\! {\rm i}\big\}$.

A more recent application of \eqref{Eqn:linear_system} for un-coded $\bf{x}$ is the massive-access scheme, where $\bf{A}$ consists of pilot signals, and $P_X(x_i)$ is jointly determined by channel coefficients and user activity \cite{Yuwei2018, Yuwei20182}. In this case, a common assumption of $P_X(x_i)$ is Bernoulli-Gaussian.  Massive access has attracted wide research interests for machine-type communications in the $5^{\rm th}$ and $6^{\rm th}$ generation (5G and 6G) cellular systems. 

Two commonly used performance measures for the system described in \eqref{Eqn:linear_system} are mean squared error (MSE) for an un-coded system and achievable rate for an FEC-coded system. The optimal limits for these measures are the minimum MSE (MMSE) and the information-theoretic capacity, respectively. For simplicity, we will say that a receiver is
\begin{itemize}
  \item MMSE-optimal if its MSE achieves MMSE for un-coded $\bf{x}$, or  
  \item capacity-optimal if its achievable rate achieves mutual information $I(\bf{x}; \bf{y})$ for coded $\bf{x}$. 
\end{itemize}
Optimal receivers under both measures often exhibit prohibitively high complexity \cite{Micciancio2001,verdu1984_1}, with a few exceptions. Some of these exceptional cases are listed below. 
\begin{itemize}
  \item The classic linear MMSE (LMMSE) detector is optimal when \eqref{Eqn:linear_system} is un-coded with IID Gaussian signaling \cite{Kay1993}.   
  \item Several compressed-sensing algorithms are asymptotically MMSE-optimal when $\bf{x}$ is un-coded and sparse, as $N\to\infty$ \cite{Donoho_CS2005, Guo_CS2009}. 
  \item When $\bf{A}$ is diagonal, \eqref{Eqn:linear_system} is equivalent to a collection of parallel single-input-single-output (SISO) sub-systems. In this scenario, Turbo \cite{Berrou1993, douillard1995iterative}, low-density parity-check (LDPC) \cite{Gallager1962, Chung01} and Polar \cite{Arikan_polar2009} codes are nearly or asymptotically capacity-optimal.
  \item Turbo-type detection algorithms \cite{Wang1999} achieve capacity-optimality when applied to properly encoded Gaussian signaling  \cite{Yuan2014, Lei20161b,YC2018TWC}. 
\end{itemize}

For discrete signaling with dense $\bf{A}$ and $\bf{x}$, however, detection methods with proven MMSE or capacity optimality, while maintaining practical complexity, remained unresolved until recently \cite{Barbier2018b, Kabashima2006, Tulino2013, Reeves_TIT2019, Barbier2017arxiv,  LeiTIT, Fan2022}. 

\subsection{AMP and Related Algorithms}\label{Sec:haar_def}
Approximate message passing (AMP) has made remarkable progress in achieving both types of optimality while maintaining practical complexity. AMP utilizes a so-called Onsager term to address the correlation issue in iterative processing \cite{Donoho2009}. A distinguished feature of AMP is its capability to precisely analyze the MSE performance through the utilization of a state-evolution (SE) technique when $M, N\to\infty$ with their ratio fixed \cite{Bayati2011, Bayati2015}. Building upon the SE methodology, the MMSE optimality of AMP was proven in \cite{Reeves_TIT2019, Barbier2017arxiv}. Furthermore, the capacity optimality of AMP is demonstrated in \cite{LeiTIT} with the assumption that the SE of coded AMP is correct and has a unique fixed point. 

Good performance of AMP is guaranteed only when $\bf{A}$ is IIDG, i.e., its entries are IID Gaussian. This IIDG restriction is relaxed in orthogonal AMP (OAMP) for a large unitarily invariant system (LUIS) \cite{Ma2016,MaSPL2015a, Ma_SPL2015b}. Let the singular value decomposition (SVD) of $\bf{A}$ be $\bf{A}\!=\!\bf{U}^{\rm H}\bf{\Sigma} \bf{V}$, where $\bf{U}\!\in\! \mathbb{C}^{M\times M}$ and $\bf{V}\!\in\! \mathbb{C}^{N\times N}$ are unitary  and $\bf{\Sigma}$ is an $M\times N$ rectangular diagonal matrix. We say that $\bf{V}$ is Haar distributed if it is uniformly distributed over all unitary matrices \cite{Hiai2000, Tulino2004}.  We say that $\bf{A}$ is \emph{right-unitarily invariant} if $\bf{V}$ is Haar distributed. We call \eqref{Eqn:linear_system} LUIS when $M,N\to\infty$ with their ratio fixed and $\bf{V}$ is Haar distributed and independent of $\bf{U}\bf{\Sigma}$. Additionally, the empirical eigenvalue distribution of $\bf{AA}^{\rm H}$ converges almost surely to a compactly supported deterministic distribution.  Recently, to avoid the high-complexity LMMSE in OAMP, two low-complexity variants called convolutional AMP (CAMP) \cite{Takeuchi2020CAMP} and memory AMP (MAMP) \cite{LeiMAMP} were proposed for LUIS. MAMP leverages a low-complexity memory matched filter to effectively suppress linear interference, making it comparable in complexity to AMP and significantly lower than that of OAMP. Notably, the dynamics of MAMP can be accurately characterized by state evolution. Furthermore, the state evolution analysis reveals that MAMP converges to the MMSE fixed point as predicted by the replica method. The MMSE and constrained capacity of LUIS can be predicted using the replica method from statistical physics \cite{Tulino2013, Kabashima2006, Ma2016}. However, the replica method involves an exchange of limits and a replica symmetry assumption, which are unproven for general LUIS. Recently, it was proven that the MMSE and constrained capacity predicted by the replica method are correct for IIDG matrices \cite{Reeves_TIT2019, Barbier2017arxiv} and certain specific sub-classes of LUIS \cite{Barbier2018b, Fan2022}. Nevertheless, a rigorous proof of the replica method for a wider range of LUIS remains an open issue.

This paper focuses on the capacity optimality of LUIS. The main technique used in this paper is the state evolution of OAMP, which is conjectured for LUIS in \cite{Ma2016} and is rigorously proved in \cite{Rangan2016, Takeuchi2017}. For convenience, we refer to the MMSE and constrained capacity predicted by the replica method as ``replica MMSE" and ``replica constrained capacity", respectively. Furthermore, we say that a receiver is MMSE-optimal if its MSE achieves the replica MMSE when $\bf{x}$ is un-coded, or capacity-optimal if its achievable rate reaches the replica constrained capacity when $\bf{x}$ is coded. In this paper, we demonstrate that OAMP can achieve the constrained capacity of LUIS provided that both the state evolution and replica methods are reliable.

\subsection{Contributions of This Paper}
This paper is dedicated to investigating the capacity optimality of OAMP. AMP-type algorithms, including OAMP, all involve iteration between two local processors: a linear estimator (LE) $\gamma$ and a non-linear estimator (NLE) $\phi$  \cite{Donoho2009,Ma2016,Rangan2016}. The performance of these two local processors can be respectively characterized by two state evolution transfer functions:  $\gamma_{\rm SE}$ and $\phi_{\rm SE}$ \cite{Bayati2011, Bayati2015, Takeuchi2017, Rangan2016}.  For more in-depth discussions on $\gamma_{\rm SE}$ and $\phi_{\rm SE}$, please refer to Subsections \ref{Sec:OAMP} and \ref{Sec:SE}. 

Following \cite{LeiTIT}, we call mutual information $I(\bf{x}; \bf{y})$ the constrained capacity of the system in \eqref{Eqn:linear_system}. It should be noted that $I(\bm{x}; \bm{y})$ solely depends on $P_X$, $\sigma^2$ and $\lambda_{\bf{AA}^{\rm H}}$ (the eigenvalues of $\bf{AA}^{\rm H}$). For each realization of $\bm{A}$ in LUIS, $\lambda_{\bf{AA}^{\rm H}}$ remains fixed as we assume the empirical eigenvalue distribution converges almost surely to a compactly supported deterministic distribution. Consequently, $I(\bm{x}; \bm{y})$ remains identical for every realization of $\bm{A}$. In \cite{LeiTIT}, we demonstrated that the achievable rate of AMP is equal to the area determined by $\gamma_{\rm SE}$ and $\phi_{\rm SE}$. When both the local NLE in AMP is MMSE-optimal, we can apply the celebrated I-MMSE theorem \cite{LeiTIT} to derive the achievable rate of AMP. We showed in \cite{LeiTIT} that this achievable rate equals the constrained capacity under matching FEC coding and accurate state evolution of coded AMP with a unique fixed point. The assumption of an MMSE-optimal NLE $\phi$ plays a crucial role in proving the constrained capacity optimality of AMP \cite{LeiTIT}. 

Unfortunately, making assumptions about the local NLE in OAMP being MMSE-optimal is not feasible. This limitation arises due to the requirement of input-output error orthogonality on the local processors. MMSE-optimal processors typically lack orthogonality, which means that the local processors in OAMP, designed to be orthogonal, are often not MMSE-optimal. This poses a significant challenge when attempting to extend the results from AMP in \cite{LeiTIT} to OAMP.

In this paper, we tackle the difficulty using a Gram-Schmidt model \cite{Schmidt1908,Yiyao_integral, LeiOAMP2022} for an orthogonal local processor. Using this model, we establish a connection between an arbitrary local processor and an orthogonal one. This connection enables us to restructure OAMP, resulting in an equivalent form where the NLE $\phi$ can be MMSE-optimal. However, the equivalent LE introduces an additional memory term as well as a complicated complex dual-input-single-output (DISO) transfer function. To address this new challenge, we elaborate a SISO variational transfer function (VTF). These locally optimal NLE and VTF allow us to leverage the I-MMSE theorem in obtaining the achievable rate of OAMP. Furthermore, we demonstrate that this achievable rate is equivalent to the replica-constrained capacity, thereby establishing the replica capacity optimality of OAMP when the state evolution of OAMP has a unique fixed point. As a special case, we specifically analyze LUIS with Gaussian signaling, in which the replica method is rigorous and the unique fixed-point condition strictly holds.
 
Due to the algorithmic equivalence between OAMP and the vector AMP (VAMP) \cite{Rangan2016}, as noted in previous works \cite{MaTWC, Takeuchi2020CAMP, LeiMAMP}, the results in this paper also apply to VAMP. Additionally, using the Gram-Schmidt orthogonalization, we will explore the similarities and differences between OAMP and the expectation propagation (EP) algorithm. We will reveal the equivalent between OAMP and EP when MMSE-optimal prototypes are used, thereby implying that EP is also replica capacity-optimal when combined with proper FEC coding. 

OAMP has garnered significant attention in various emerging applications, including MIMO channel estimation and massive access \cite{MaTWC, Yiyao_mmv, Khani2020, Zhang2019, Yiyao_ofdm}, as it offers versatility beyond Gaussian signaling, sparsity, and IIDG sensing matrices. While most existing works on OAMP focus on un-coded scenarios, the findings in this paper present an optimization technique for coded cases. Notably, these results have recently been extended to generalized multiuser MIMO (GMU-MIMO) communications \cite{Yuhao_tcom}. By leveraging the capacity-area theorem presented in this paper \cite{Yuhao_tcom}, the constrained capacity region of GMU-MIMO is established. Moreover, the optimal multi-user coding scheme for OAMP, derived using the matching principle, is proposed to achieve the constrained capacity region of GMU-MIMO \cite{Yuhao_tcom}.  Furthermore, building upon the results in this paper, the capacity optimality investigation of MAMP has been recently undertaken in \cite{Code_MAMP}. Along with these, the findings of this paper were extended to the generalized linear model (GLM) \cite{Lei_goamp}, where the generalized OAMP (GOAMP) comprises a dual-input-dual-output linear detector paired with two nonlinear detectors, leading to intricate achievable rate analysis. To tackle this challenge, an equivalent SISO variational state evolution is constructed, facilitating the extension of the capacity-area theorem and matching principle in this paper. This enables an accurate evaluation of the maximum achievable rate of GOAMP \cite{Lei_goamp}.

\subsection{Main Results}
This paper aims to address the following fundamental questions concerning LUIS and OAMP:
\begin{enumerate}[Q1.] 
    \item What is the relationship between the constrained capacity of LUIS and the transfer functions of OAMP, i.e., particularly in terms of the area-capacity property of LUIS? 
    \item What is the optimal coding principle for the OAMP receiver? 
    \item What is the maximum achievable rate of the OAMP receiver? Can OAMP achieve the constrained capacity of LUIS?
    \item How much rate loss occurs due to factors such as non-Gaussian signaling, SISO channel coding, non-iterative LMMSE detection, cross-symbol interference, and channel noise? 
\end{enumerate}

\begin{figure}[t] %
  \centering
  \includegraphics[width=5.5cm]{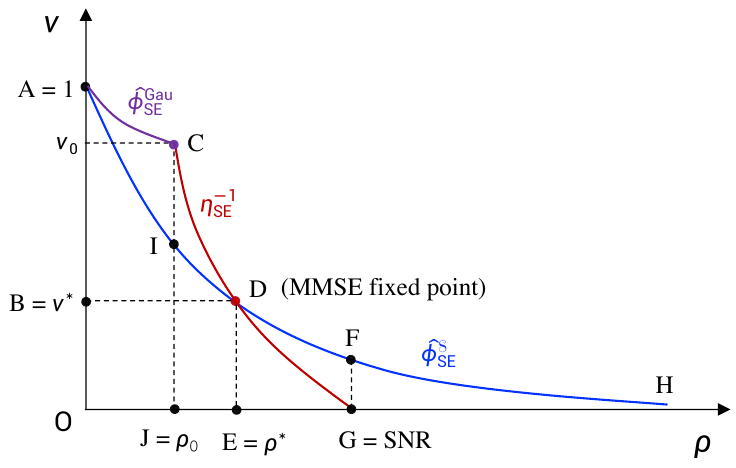}\\ %
  \caption{Graphic illustrations of the VTF in \eqref{Eqn:TF} for OAMP detection and the area properties of LUIS. $(\rho^*, v^*)$ is the fixed point of \eqref{Eqn:TF}. ${\rm H} = (0, \infty)$, $\hat{\phi}_{\rm SE}^{\rm Gau}(\rho)=1/(1+\rho)$, $\rho_0=[\gamma_{\rm SE}(1)]^{-1}-1$, $v_0=\gamma^{-1}_{\rm SE}(\rho_0)$ and $\eta_{\rm SE}(0)={\rm SNR}$. $A_{\rm ADGO}$: replica constrained capacity,  $A_{\rm ACD}$: shaping gain of Gaussian signaling, $A_{\rm ADEO}$: achievable rate of the cascading (iterative) OAMP receiver, $A_{\rm AIJO}$: achievable rate of the cascading (non-iterative) LMMSE receiver, $A_{\rm DGE}$: rate loss of the cascading OAMP receiver, $A_{\rm DGE}$: rate loss caused by cross-symbol interference, $A_{\rm DGE}$: rate loss caused by the channel noise.} \label{Fig:TF_chart}  
\end{figure}

\begin{figure}[t] %
\centering 
\includegraphics[width=5.5cm]{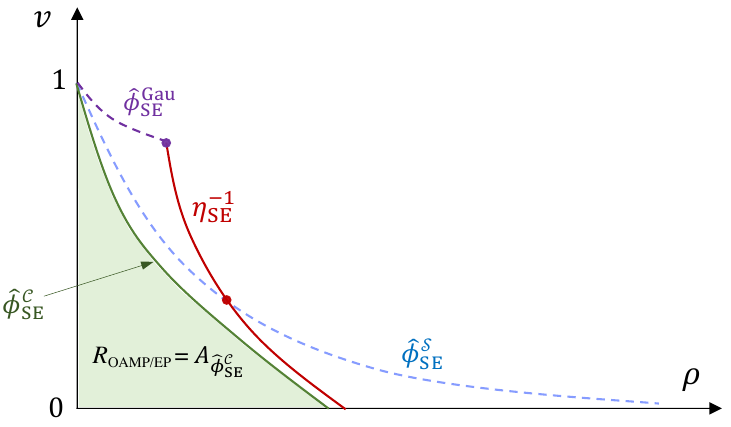}\\  %
\caption{Graphic illustration of transfer curves $\tilde{\eta}_{\rm SE}^{-1}$, $\hat{\phi}_{\rm SE}^{\mathcal{S}}$ and $\hat{\psi}_{\rm SE}^{\mathcal{C}}$ in OAMP. $A_{\hat{\phi}_{\rm SE}^{\mathcal{C}}}$ (the area covered by $\hat{\psi}_{\rm SE}^{\mathcal{C}}$) represents the achievable rate of the OAMP/EP receiver. $\hat{\phi}_{\rm SE}^{\mathcal{S}}$ denotes demodulation transfer function in un-coded case and $\hat{\phi}_{\rm SE}^{\mathcal{C}}$ denotes decoding transfer function in coded case. $ \hat{\phi}_{\rm SE}^{\cal C}(\rho) \leq  \hat{\phi}_{\rm SE}^{\cal S}(\rho)$ follows the decoding gain, and $\hat{\phi}_{\rm SE}^{\cal C}(\rho) <  {\eta}_{\rm SE}^{-1}({\rho})$ follows the error-free condition.}\label{Fig:track}%
\end{figure}

We obtain the following results in light of the aforementioned questions.

\subsubsection{Capacity-Area Theorem} Given the variational LE and NLE transfer functions $\eta_{\rm SE}$ and $\phi_{\rm SE}^{\cal S}$ (as defined in  \eqref{Eqn:TF}) of the un-coded OAMP, this paper, as illustrated in Figure \ref{Fig:TF_chart}, proves that the replica constrained capacity of LUIS can be represented by the area $A_{\rm ADGO}$ bounded by ${\eta}_{\rm SE}^{-1}$ (the inverse of $\eta_{\rm SE}$) and $\hat{\phi}_{\rm SE}^{\cal S}$, i.e., 
\BE \label{Eqn:C_area}
     C_{\rm Rep}  = A_{\rm ADGO}, 
\EE
when the state evolution of OAMP has a unique fixed point. For the details, refer to Theorem \ref{The:area_LUIS} and its discussions in Section \ref{Sec:rate_OAMP}.  This finding answers the first question and offers essential guidance for coding optimization, as well as for proving the capacity optimality of the OAMP receiver.
 
\subsubsection{Coding Principle for OAMP Receiver} Let $\phi_{\rm SE}^{\cal C}$ (see  \eqref{Eqn:SE_equiv_Cb}) denote the Lipschitz continuous MMSE transfer function of the APP decoder. As shown in Fig. \ref{Fig:track}, the achievable rate of LUIS with joint OAMP detection and decoding is determined by the area $A_{\hat{\phi}_{\rm SE}^{\cal C}}$ covered by $\hat{\phi}_{\rm SE}^{\cal C}$:
\BE
    R_{\rm OAMP} = A_{\hat{\phi}_{\rm SE}^{\cal C}},\label{Eqn:Rate_OAMP}
\EE
under the conditions that:
\begin{itemize} 
    \item  $\hat{\phi}_{\rm SE}^{\cal C}\leq \hat{\phi}_{\rm SE}^{\cal S}$ due to the decoding gain since the decoding MSE should be not worse than the demodulation.
    \item $\hat{\phi}_{\rm SE}^{\cal C}< {\eta}_{\rm SE}^{-1}$, indicating that there should be a tunnel between $\phi_{\rm SE}^{\cal C}$  and ${\eta}_{\rm SE}^{-1}$ to ensure error-free recovery.  
\end{itemize}
Therefore, the optimal coding principle is matching the decoding transfer function $\hat{\phi}_{\rm SE}^{\cal C}$ with $\min\{{\eta}_{\rm SE}^{-1},\; \hat{\phi}_{\rm SE}^{\cal S}\}$, i.e., 
\BE
    \hat{\phi}_{\rm SE}^{\cal C}\to \min\{{\eta}_{\rm SE}^{-1}({\rho}),\; \hat{\phi}_{\rm SE}^{\cal S}(\rho)\},\label{Eqn:matching_curve}
\EE
which maximizes the achievable rate of LUIS with the OAMP receiver. For the details, refer to Lemma \ref{Lem:Rate_VTF} and its discussions in Subsections \ref{Sec:VT_cod}-\ref{Sec:area_AMP}.

\subsubsection{Replica Constrained Capacity Optimality} Following \eqref{Eqn:Rate_OAMP} and \eqref{Eqn:matching_curve}, the maximum achievable rate of the OAMP receiver is equal to $A_{\rm ADGO}$, i.e., $ R_{\rm OAMP}\to A_{\rm ADGO}$. Furthermore, following the capacity-area theorem in \eqref{Eqn:C_area}, when $\hat{\phi}_{\rm SE}^{\cal C}\to \min\{{\eta}_{\rm SE}^{-1}({\rho}),\; \hat{\phi}_{\rm SE}^{\cal S}(\rho)\}$, we have   
\BE
    R_{\rm OAMP}\to C_{\rm Rep}, 
\EE
i.e., OAMP achieves the replica constrained capacity optimality of LUIS when the state evolution of OAMP has a unique fixed point. For the details, refer to Theorem \ref{The:cap_opt} and its discussions in Subsection \ref{Sec:area_AMP}. 

\subsubsection{Area Properties} As shown in Fig. \ref{Fig:TF_chart}, we also present some intriguing area properties of LUIS relevant to the rate losses caused by non-Gaussian signaling (i.e., area $A_{\rm ACD}$), SISO channel coding (i.e., area $A_{\rm DGE}$), non-iterative LMMSE detection (i.e., area $A_{\rm IDEJ}$), cross-symbol interference (i.e., area $A_{\rm DFG}$), and channel noise (i.e., area $A_{\rm FHG}$). For the details, refer to Subsection \ref{Sec:areas}.

\subsection{Notations}
Boldface lowercase letters represent vectors and boldface uppercase symbols denote matrices. We say that $x=x_{\rm Re}+ {\rm i}\cdot x_{\rm Im}$ is circularly-symmetric complex Gaussian (CSCG) if $x_{\rm Re}$ and $x_{\rm Im}$ are two independent Gaussian distributed random variables with ${\rm E}\{x_{\rm Re}\} = {\rm E}\{x_{\rm Im}\} = 0$ and $ {\rm Var}\{x_{\rm Re}\} = {\rm Var}\{x_{\rm Im}\}$. We define ${\rm Var}\{x\} \equiv {\rm Var}\{x_{\rm Re}\} + {\rm Var}\{x_{\rm Im}\}$. Denote $I({\bf{\bf{x}};{\bf{y}}})$ for the mutual information between $\bf{x}$ and $\bf{y}$, $\mb{I}$ for the identity matrix with a proper size, $\bm{a}^{\mr{H}}$ for the conjugate transpose of $\bm{a}$, $\|\bm{a}\|$ for the $\ell_2$-norm of the vector $\bm{a}$, $\det(\bf{A})$ for the determinant of $\bf{A}$, $\ms{tr}(\bm{A})$ for the trace of $\bm{A}$, $A_{ij}$ for the $i$th-row and $j$th-column element of $\bf{A}$, $\mathcal{CN}(\bm{\mu},\bm{\Sigma})$ for the CSCG distribution with mean $\bm{\mu}$ and covariance $\bm{\Sigma}$, $\ms{E}\{\cdot\}$ for the expectation operation over all random variables involved in the brackets, unless otherwise specified. $\mr{E}\{a|b\}$ for the expectation of $a$ conditional on $b$, and  $\ms{var}\{{a}\}$ for $\ms{E}\left\{ \|{a} - \ms{E}\{{a}\}\|^2 \right\}$, $\ms{mmse}\{{a}|{b}\}$ for $\ms{E}\left\{ \|{a} - \ms{E}\{{a}|{b}\}\|^2|{b} \right\}$. 

Throughout this paper, unless stated otherwise, we will assume that (i) the length of a vector is $N$, (ii) $\bf{x}$ is normalized, i.e., $\frac{1}{N}\mr{E}\{\|{\bf{x}}\|^2\}=1$, and (iii) $\bf{A}$ is normalized, i.e., $\frac{1}{N}\mr{tr}\{\bf{A}^{\rm H}\bf{A} \}=1$. In this paper, we do not develop new notations to distinguish between random variables and their representations, or deterministic variables since they can be easily distinguished by context. For a fixed input distribution $P_X(x)$, the mutual information denotes the constrained capacity. For convenience,  the constrained capacity is called ``capacity" when it is clear from the context.

\section{System Model and Preliminaries}%

A large unitarily invariant system (LUIS) in \eqref{Eqn:linear_system} consists of a linear constraint $\Gamma$ and a non-linear constraint $\Phi$:
\BS\begin{align}
{\rm Linear\; constraint}\;\; \Gamma:\quad &\bf{y}=\bf{Ax}+\bf{n},  \label{Eqn:LC}\\
{\rm Non\!\!-\!\!linear\; constraint}\;\; \Phi:\quad  &\bf{x}\sim P_{\bf{X}}(\bf{x}).\label{Eqn:NLC} 
\end{align}\ES
The LUIS is assumed to meet the following assumption. \vspace{1mm} 

\begin{assumption}\label{ASS:Model}
 We assume that $\bf{A}$ is a \emph{right-unitarily invariant} measurement matrix (see \ref{Sec:haar_def}) and the average power of  $\bf{x}$ is normalized, i.e., $\frac{1}{N}\|{\bf{x}}\|^2=1$, and ${\rm SNR} = \sigma^{-2}$ is the transmit signal noise ratio (SNR). We consider a large-scale LUIS that $M,N\to\infty$ with a fixed $\beta=N/M$. In addition, the empirical eigenvalue distribution of $\bf{AA}^{\rm H}$ converges almost surely to a deterministic distribution with compact support in the large system limit. Furthermore, we assume that only the receiver\footnote{This assumption has been commonly used in MIMO and/or multi-user communications\cite{MaTWC, Lei20161b}. If the transmitter also has $\bf{A}$,  the LUIS can be converted to parallel SISO  channels. Then, water filling is capacity optimal.} knows $\bf{A}$. 
\end{assumption}

\subsection{Un-coded LUIS}\label{Sec:uncod_LUIS}
{For an un-coded LUIS,  
the non-linear constraint in \eqref{Eqn:NLC} can be rewritten in a symbol-by-symbol manner as:
\BE\label{Eqn:Phi_S}
    {\rm Constellation\; constraint}\;\; \Phi_{\cal S}:\quad x_i\sim P_X(x),\; \forall i.
\EE
That is, the entries of $\bf{x}$ are IID with distribution $P_X(x)$.  }

In an un-coded LUIS, the MSE serves as a widely used performance metric. For Gaussian distribution $P_X(x)$, the optimal solution is given by the standard LMMSE estimate. However, for non-Gaussian distribution $P_X(x)$, finding the optimal solution is generally NP-hard \cite{Micciancio2001,verdu1984_1}.  
   
\emph{Minimum Mean Square Error (MMSE):} The goal in un-coded LUIS  is to find an MMSE estimation of $\bf{x}$. That is, the estimation MSE converges to
\BE\label{Eqn:post_mean}
{\rm mmse}\{\bf{x}|\bf{y}, \bf{A}, {\Gamma}, \Phi \} \equiv \tfrac{1}{N}{\mr{E}}\{\|\hat{\bf{x}}_{\rm post}-{\bf{x}}\|^2\},
\EE
where $\hat{\bf{x}}_{\rm post}\!=\!{\mr{E}}\{\bf{x}|\bf{y}, \bf{A}, {\Gamma}, \Phi \}$ is the \textit{a-posteriori} mean of $\bf{x}$.

The asymptotic MMSE in the large system limit is typically predicted using the \emph{replica method}, which is summarized in Appendix \ref{APP:replica}.

\subsection{Coded LUIS} In the un-coded LUIS, an error-free recovery can not be guaranteed. To achieve an error-free estimation, we consider the LUIS with forward error control (FEC) coding. In a coded LUIS, the non-linear constraint in \eqref{Eqn:NLC} is modified to:
\BE\label{Eqn:Phi_C}
   {\rm Code\; constraint}\;\; \Phi_{\cal C}:\quad \bf{x} \in \bf{\mathcal{C}} \;\; {\rm and} \;\; x_i\sim P_X(x),\; \forall i.
\EE

In the coded LUIS, the achievable rate is commonly used as a performance metric. For a specific receiver, the code rate $R_{\bf{\mathcal{C}}}$ of codebook $\bf{\mathcal{C}}$ is achievable if its estimation tends to be error-free in the asymptotic sense, i.e., ${\rm mmse}\{\bf{x}|\bf{y}, \bf{A}, {\Gamma}, \Phi_{\cal C} \}\equiv \tfrac{1}{N}{\mr{E}}\{\|{\rm E}\{\bf{x}|\bf{y}, \bf{A}, {\Gamma}, \Phi_{\cal C} \}-{\bf{x}}\|^2\}\to 0$ or equivalently, the block error probability$\to 0$. A natural upper bound on the achievable rate is given by the constrained capacity of LUIS, i.e., the mutual information given $x\sim P_X(x)$: $R_{\bf{\mathcal{C}}}\le C_{\rm LUIS}=I(\bf{x};\bf{y})$.
 
Note that the eigenvalues of $\bf{AA}^{\rm H}$ are fixed in LUIS. Therefore, $I(\bm{x}; \bm{y})$ remains constant for each realization of $\bm{A}$. This implies that the constrained capacity of LUIS $C_{\rm LUIS}$ can be calculated using any realization of $\bm{A}$ since it does not depend on the specific realization. Given this, our aim is to develop a coding scheme and a receiver with practical complexity  that can achieve a rate as close as possible to the constrained capacity of LUIS, i.e., $ R_{\bf{\mathcal{C}}}\to C_{\rm LUIS}$.

For Gaussian signaling, $C_{\rm LUIS}$ is the well-known Gaussian capacity \cite{David2005} below. 

\begin{lemma} 
The Gaussian capacity of LUIS, assuming $\bf{x}\sim\mathcal{CN}(\bf{0},\bf{I})$, is expressed as 
\BE \label{Eqn:Gau_C} 
C_{\rm Gau}({\rm SNR})
= \tfrac{1}{N}\log \det\big(\bf{I} +  {\rm SNR}\bf{A}^H\bf{A}\big).
\EE 
\end{lemma} 

Determining the constrained capacity of LUIS for arbitrary signal distributions, which can be non-Gaussian, is nontrivial and was conjectured using the replica method (see Appendix \ref{APP:replica}) \cite{Tulino2013, Kabashima2006}. However, designing a coding scheme and receiver with practical complexity that can achieve the replica constrained capacity remains an open issue. In this paper, we will prove the replica capacity optimality of OAMP based on matched FEC coding.

\subsection{Orthogonal Approximate Message Passing (OAMP)} 
The OAMP algorithm consists of two orthogonal estimators. The concept of orthogonal estimators was initially introduced in \cite{Ma2016}.  In this section, we outline a general approach to construct orthogonal estimators using Gram-Schmidt  orthogonalization (GSO), which includes the de-correlated linear estimator \cite{Ma2016}, divergence-free (e.g., differential-based) estimator \cite{Ma2016, Donoho2009, Rangan2016}, integral-based orthogonal estimator \cite{Yiyao_integral}, and expectation propagation (EP) \cite{Cakmak2018, opper2005expectation, Minka2001} as special instances. In essence, the GSO method offers greater generality compared to existing literature. For example, conventional methods like AMP \cite{Donoho2009}, OAMP  \cite{Ma2016} and VAMP  \cite{Rangan2016} are constrained to differentiable estimators due to the use of divergence operation. In contrast, GSO-based OAMP overcomes this limitation and can accommodate non-differentiable estimators. 

\subsubsection{Gram-Schmidt (GS) Model}  
Let ${\rm E}\{\|\bf{x}\|^2\}/N=1$. Then, we can express an arbitrary observation  $\hat{\bf{x}}$ as a Gram-Schmidt (GS) model with respect to $\bf{x}$ \cite{Schmidt1908,Yiyao_integral, LeiOAMP2022}:
\BE\label{Eqn:GP_model}
 \hat{\bf{x}} = \alpha \bf{x} + \bf{\xi},
\EE 
where $\alpha=\tfrac{1}{N}\mr{E}\{\hat{\bf{x}}^{\mr{H}}{\bf{x}}\}$. Its average entry-wise power is $v=\tfrac{1}{N}\mr{E}\{\|\bf{\xi}\|^2\}=\tfrac{1}{N}\mr{E}\{\bf{\xi}^\mr{H}\bf{\xi} \}$. It can be verified that $\bf{\xi}$ is orthogonal to $\bf{x}$, i.e.,
 \BE\label{Eqn:GP_modelc}
\mr{E}\{\bf{x}^{\mr{H}} \bf{\xi}\} = 0.
\EE
The error term $\bf{\xi}$ in \eqref{Eqn:GP_model} has a distinct definition compared to the conventional error $\hat{\bf{x}}-\bf{x}$. For clarity,  we refer to $\bf{\xi}$ as the GS error of $\hat{\bf{x}}$, and $\{\alpha, v\}$ as the GS parameters. Assuming a given distribution of $\bf{x}$ and that $\bf{\xi}$ consists of IIDG entries with zero mean, the distribution of $\hat{\bf{x}}$ in \eqref{Eqn:GP_model} is determined by the GS parameters ($\alpha$ and $v$). 
 
\subsubsection{Orthogonal Estimator} 
Orthogonal estimators are building blocks in OAMP. Consider an estimator of $\bf{x}$: $\bf{x}_{\rm out} = f(\bf{x}_{\rm in})$, 
where $\bf{x}_{\rm in}$ and $\bf{x}_{\rm out}$ can be expressed in their respective GS models: 
\BE\label{Eqn:GSO_f} 
 \bf{x}_{\rm in}  = \alpha_{\mr{in}}\bf{x}+\bf{\xi}_{\mr{in}}  \; {\rm and} \; 
   \bf{x}_{\rm out}  = \alpha_{\mr{out}}\bf{x}+{\bf{\xi}}_{\mr{out}}.
\EE
\begin{definition}[Orthogonal Estimator]
 We say that $f(\cdot)$ is an orthogonal estimator if
\BE 
    {\rm E}\big\{\bf{\xi}_{\mr{in}}^{\rm H} {\bf{\xi}}_{\mr{out}}\big\}=0.\label{Eqn:Orthogonality_fa}
\EE
Since ${\rm E}\{\bf{\xi}_{\mr{in}}^{\rm H}\bf{x}\}\!=\!0$,  \eqref{Eqn:Orthogonality_fa} is equivalent to $\mr{E}\big\{\bf{\xi}_{\mr{in}} ^{\mr{H}}f(\bf{x}_{\mr{in}})\big\}\!=\!0$.  
\end{definition}

\subsubsection{Orthogonal Estimator Construction} \label{Sec:GSO}
 The initial development of specific orthogonal estimators from given prototypes was presented in \cite{Ma2016}, using differentiable functions. An integral approach was recently introduced in \cite{Yiyao_integral, Yiyao_mmv} using GSO \cite{Schmidt1908}. Notably, GSO does not require differentiability and hence is more general. The discussions in this paper will be based on GSO. 

\begin{figure}[t] 
  \centering
  \includegraphics[width=2.8cm]{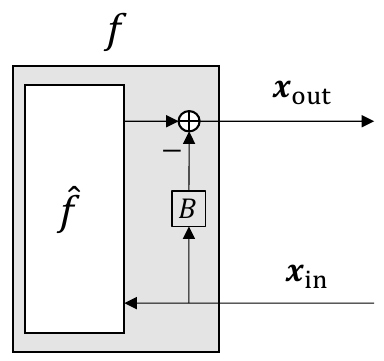}\\
  \caption{Construction of an orthogonal estimator $f(\cdot)$ via the GSO in \eqref{Eqn:f_orth} based on an arbitrary prototype $\hat{f}(\cdot)$, which may be non-orthogonal. $B$ is the GSO coefficient given in \eqref{Eqn:GSO_B}. Consequently, the GS error orthogonality ${\rm E}\big\{\bf{\xi}_{\mr{in}}^{\rm H} {\bf{\xi}}_{\mr{out}}\big\}=0$ holds for $f(\cdot)$, where $\bf{\xi}_{\mr{in}}$ and $\bf{\xi}_{\mr{out}}$ are the GS errors of $\bf{x}_{\mr{in}}$ and $\bf{x}_{\mr{out}}$, respectively.}\label{Fig:Orth_f}  
\end{figure}
 
\emph{GS Orthogonalization (GSO)}: Consider an arbitrary prototype $\hat{f}(\cdot)$, which is typically not an orthogonal estimator. In this part, we explore techniques to realize the desired orthogonality stated in \eqref{Eqn:Orthogonality_fa}. As illustrated in Fig.~\ref{Fig:Orth_f}, we construct an orthogonal $f(\bf{x}_{\mr{in}})$ by
\BE\label{Eqn:f_orth}
\bf{x}_{\mr{out}}=f(\bf{x}_{\mr{in}})=\hat{f}(\bf{x}_{\mr{in}})- B \bf{x}_{\mr{in}}.
\EE 
Then, the orthogonality requirement can be expressed as
\begin{align}\label{Eqn:GSO_ort}
\mr{E}\big\{\bf{\xi}_{\mr{in}}^{\mr{H}}\bf{\xi}_{\rm out}\big\} &=  \mr{E}\big\{\bf{\xi}_{\mr{in}}^{\mr{H}}[f(\bf{x}_{\mr{in}})-\alpha_{\rm out}\bf{x}]\big\}\\
&\!\mathop  = \limits^{({\rm{a}})} \! \mr{E}\big\{\bf{\xi}_{\mr{in}} ^{\mr{H}}f(\bf{x}_{\mr{in}})\big\}\\
& \!\mathop  = \limits^{({\rm{b}})}   \mr{E}\left\{\bf{\xi}_{\mr{in}} ^{\mr{H}}[\hat{f}(\bf{x}_{\mr{in}})- B \bf{x}_{\mr{in}}]\right\}\\
& = 0,
\end{align} 
where equality (a) is due to ${\rm E}\big\{\bf{x}^{\rm H} \bf{\xi}_{\rm in}\big\}=0$, and (b) due to \eqref{Eqn:f_orth}. Noting that $\mr{E}\big\{ \bf{\xi}_{\mr{in}}^{\mr{H}}\bf{x}_{\mr{in}}\big\} =\mr{E}\big\{\bf{\xi}_{\mr{in}}^{\mr{H}}{(\alpha_{\mr{in}}\bf{x}+\bf{\xi}_{\mr{in}})}\big\} =\mr{E}\big\{\|{\bf{\xi}}_{\mr{in}}\|^2\big\}$ (recalling  ${\rm E}\big\{\bf{x}^{\rm H} \bf{\xi}_{\rm in}\big\}=0$), from \eqref{Eqn:GSO_ort} we have
\BE\label{Eqn:GSO_B}
B= \mr{E}\{\bf{\xi}_{\mr{in}}^{\mr{H}}\hat{f}(\bf{x}_{\mr{in}})\}/\mr{E}\{\|{\bf{\xi}}_{\mr{in}}\|^2\}.
\EE

Assume that $\bf{x}_{\rm in} = \bf{x}+ \bf{\xi_{\mr{in}}}$ {with $\bf{\xi_{\mr{in}}}\sim \mathcal{CN}(\bf{0},v_{\rm in}\bf{I})$} and $\hat{f}(\cdot)$ achieves local MMSE, i.e., $\hat{f}(\bf{x}_{\rm in} )={\rm E}\{\bf{x}|\bf{x}_{\rm in} \}$. Then 
\BE\label{Eqn:MMSE_df}
B  = v_{\hat{f}}/v_{\mr{in}},
\EE
where $v_{\hat{f}}\equiv \frac{1}{N}\mr{E}\big\{\| \hat{f}({\bf{x}}_{\mr{in}}) - \bf{x}\|^2\big\}$ and $v_{\mr{in}}\equiv \frac{1}{N}\mr{E}\big\{\|{\bf{\xi}}_{\mr{in}}\|^2\big\}$. See Appendix \ref{APP:Pro_EP} for the proof of \eqref{Eqn:MMSE_df}.

\emph{Connection to Expectation propagation (EP)}: EP is a heuristic method that relies on the Gaussian assumption for the input and output messages. Interestingly, the GSO in \eqref{Eqn:MMSE_df} aligns with the EP updating rule in \cite{Cakmak2018, opper2005expectation, Minka2001}. That is, EP and GSO are equivalent when locally optimal prototypes are employed. Otherwise, they are not. In other words, the local estimators in EP exhibit orthogonality only when they are locally MMSE optimal. Overall, GSO is a more versatile method to construct orthogonal local estimators that can be optimal or sub-optimal, linear or non-linear. Therefore, this paper also establishes the replica capacity optimality of EP under the same assumptions as those for OAMP such as the reliability of state evolution and the replica method. 

\emph{Connection to the Original OAMP/VAMP}: The original OAMP \cite{Ma2016} and VAMP \cite{Rangan2016} both rely on the assumption of a differentiable $\hat{f}(\cdot)$. However, as demonstrated in Appendix \ref{APP:Pro_EP}, differentiability is not necessary for the orthogonality with MMSE optimal prototypes. This implies that the replica capacity optimality of OAMP demonstrated in this paper is not restricted solely to differentiable local estimators.

\subsubsection{Orthogonal Approximate Message Passing (OAMP)}\label{Sec:OAMP}

\begin{figure}[t] 
  \centering
  \includegraphics[width=5.5cm]{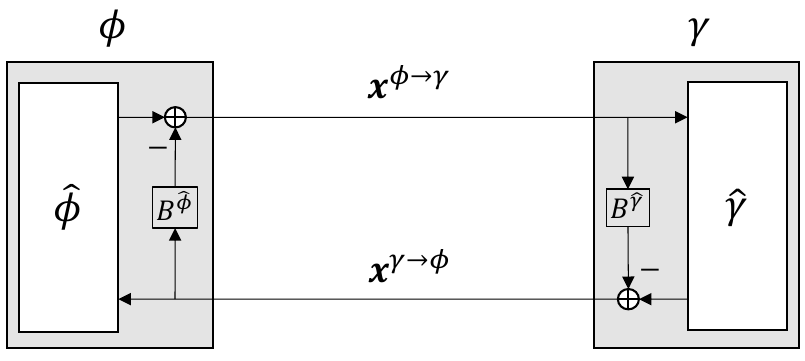}\\  
  \caption{Graphical illustration of OAMP receiver involving two local orthogonal estimators $\gamma$ (for $\Gamma$) and $\phi$ (for $\Phi$), which are constructed by GSO in \eqref{Eqn:OAMP} based on the prototypes $\hat{\gamma}$ and $\hat{\phi}$, respectively. $B^{\hat{\gamma}}$ and $B^{\hat{\phi}}$ are GSO coefficients given in \eqref{Eqn:GSO_B}. Hence, the orthogonalities ${\rm E}\big\{\big(\bf{\xi}^{\gamma\to \phi}\big)^{\rm H} \bf{\xi}^{\phi\to \gamma}\big\}=0$ and ${\rm E}\big\{\big(\bf{\xi}^{\phi\to \gamma}\big)^{\rm H} \bf{\xi}^{\gamma\to \phi}\big\}=0$ hold for $\gamma$ and $\phi$, where $\bf{\xi}^{\gamma\to \phi}$ and $\bf{\xi}^{\gamma\to \phi}$ are the GS errors of $\bf{x}^{\gamma\to \phi}$ and $\bf{x}^{\gamma\to \phi}$, respectively. For simplicity, the iterative index $t$ is omitted in the figure.}\label{Fig:OAMP} 
\end{figure}

We are now ready to define OAMP formally. We add an iteration index $t$ and express the iterative process in Fig.~\ref{Fig:OAMP} as follows.

\textit{Generic Iterative Process (GIP):} Initializing from $t=0$ and $ \bf{x}^{\phi\to \gamma}_0= \bf{0}$, 
\BS\label{Eqn:Ite_Proc_para}\begin{alignat}{2}
{\rm Linear \; estimator\;(LE):} \;\; \bf{x}^{\gamma\to \phi}_t & = \gamma_t\big(\bf{x}^{\phi\to \gamma}_{t}\big), \\
{\rm Non\!\!-\!\!linear \; estimator\;(NLE):} \;\; \bf{x}^{\phi\to \gamma}_{t+1} & = \phi_t\big(\bf{x}^{\gamma\to \phi}_t \big).
\end{alignat}\ES 
 
In \eqref{Eqn:Ite_Proc_para}, $\gamma_t(\cdot)$ and $\phi_t(\cdot)$
respectively generate refined estimates of $\bf{x}$. Proper statistical models (e.g., GS models) of $\bf{x}^{\phi\to \gamma}_{t}$ and $\bf{x}^{\gamma\to \phi}_t$ are required for the effective design of $\gamma_t(\cdot)$ and $\phi_t(\cdot)$, respectively. Tracking such models during iterative processing is in general a prohibitively difficult task. This difficulty is resolved by OAMP using an orthogonal principle.

Let the messages in \eqref{Eqn:Ite_Proc_para} be expressed in their GS models: 
\BE\label{Eqn:GSO_gamma_phi}  
  \bf{x}^{\phi\to \gamma}_{t} = \alpha_{t}^{\phi\to \gamma}\bf{x}+\bf{\xi}_{t}^{\phi\to \gamma} \; {\rm and} \; \bf{x}^{\gamma\to \phi}_t   = \alpha_{t}^{\gamma\to \phi}\bf{x}+{\bf{\xi}}^{\gamma\to \phi}_{t}.
 \EE
Let the average powers of the GS errors $\bf{\xi}_{t}^{\phi\to \gamma}$ and ${\bf{\xi}}^{\gamma\to \phi}_{t}$ be $v_t^{\phi\to \gamma}$ and $v^{\gamma\to \phi}_{t}$, respectively.

 \textit{Orthogonal AMP (OAMP):} A GIP in \eqref{Eqn:Ite_Proc_para} is referred to as OAMP when the following orthogonal constraints hold for $N\to \infty$, $t\ge 0$, 
\BE\label{Eqn:Orthogonality_oamp} 
 \!\!\! {\rm E}\Big\{\big(\bf{\xi}_{t}^{\gamma\to \phi}\big)^{\rm H} \bf{\xi}_{t}^{\phi\to \gamma}\Big\}=0  \; {\rm and} \;
{\rm E}\Big\{\big(\bf{\xi}_{t}^{\phi\to \gamma}\big)^{\rm H} \bf{\xi}_{t+1}^{\gamma\to \phi}\Big\}=0. 
\EE
That is, $\gamma_t$ and $\phi_t$ in \eqref{Eqn:Ite_Proc_para} are orthogonal estimators. 

The orthogonality in \eqref{Eqn:Orthogonality_oamp} plays a crucial role in addressing the correlation problem. The GSO  in Section \ref{Sec:GSO} can be employed to construct the orthogonal estimators in OAMP. Specifically, let $\hat{\gamma}_t$ and $\hat{\phi}_t$ be the prototypes of LE and NLE, respectively. Then, OAMP can be constructed as
\BS\label{Eqn:OAMP}\begin{align}
      \bf{x}^{\gamma\to \phi}_t & = {\gamma}_t\big(\bf{x}^{\phi\to \gamma}_{t}\big)= \hat{\gamma}_t\big(\bf{x}^{\phi\to \gamma}_{t}\big)- B_t^{\hat{\gamma}} \bf{x}^{\phi\to \gamma}_{t}, \\
      \bf{x}^{\phi\to \gamma}_{t+1} &= {\phi}_t\big(\bf{x}^{\gamma\to \phi}_t \big) = \hat{\phi}_t\big(\bf{x}^{\gamma\to \phi}_t \big)- B_t^{\hat{\phi}} \bf{x}^{\gamma\to \phi}_{t},
\end{align}\ES 
where $B^{\hat{\gamma}}_t$ and $B^{\hat{\phi}}_t$ are GSO coefficients given in \eqref{Eqn:GSO_B}. Fig.~\ref{Fig:OAMP} illustrates the block diagram of OAMP. In this paper, we focus on locally MMSE optimal prototypes: $ \hat{\gamma}_t\big(\bf{x}^{\phi\to \gamma}_{t}\big) ={\rm E}\{\bf{x}|\bf{x}^{\phi\to \gamma}_{t},\Gamma\}$ and $\hat{\phi}_t\big(\bf{x}^{\gamma\to \phi}_t \big)  ={\rm E}\{\bf{x}|\bf{x}^{\gamma\to \phi}_t,\Phi\}$. In this case, EP and OAMP are equivalent.

\emph{Discussions:} The orthogonalization plays a central role in OAMP, as it guarantees the error Gaussianity of an iterative process \cite{Ma2016, Takeuchi2017}. This Gaussianity enables us to accurately characterize the iterative process using transfer functions. In fact, the performance of the orthogonalized estimators (e.g., $\gamma_t$ and $\phi_t$) may be locally inferior to that of the un-orthogonalized estimators (e.g., $\hat{\gamma}_t$ and $\hat{\phi}_t$). However, we will demonstrate that the orthogonalized iterative process converges to a globally optimal solution, such as the replica capacity or MMSE \cite{Barbier2018b, Tulino2013, Kabashima2006}, for coded or uncoded LUIS. On the contrary, if we directly use un-orthogonalized estimators for an iterative process, we may obtain better local transfer functions. However, due to the correlation problem, the iterative process cannot be accurately characterized by these local transfer functions. Consequently, its global performance is generally worse than the orthogonalized iterative process.

\subsubsection{State Evolution (SE) of OAMP}\label{Sec:SE_OAMP}
Define the true errors as
\BE\label{Eqn:errors}
\bf{e}_{t}^{{\phi}\to\gamma}  \equiv  \bf{x}_{t}^{{\phi}\to\gamma}  -\bf{x} \; {\rm and} \;  \bf{e}_{t}^{{\gamma}\to\phi}  \equiv  \bf{x}_{t}^{{\gamma}\to\phi}  -\bf{x}.
\EE 
 Let $\vartheta_t$ be the MSE of $\bf{x}_{t}^{{\phi}\to\gamma}$ and $\rho_t$ the signal to interference plus noise ratio (SINR) of $\bf{x}_{t}^{\gamma\to{\phi}}$: 
 \BE
  \vartheta_t   \equiv  \tfrac{1}{N} {\rm E}\big\{\| \bf{e}_{t}^{{\phi}\to\gamma} \|^2\big\}  \; {\rm and} \; 
  \rho_t   \equiv N/ {\rm E}\big\{\| \bf{e}_{t}^{\gamma\to{\phi}} \|^2\}.  
\EE

\begin{assumption}\label{ASS:phi}
  We assume that the NLE  $\hat{\phi}_t$ in OAMP is Lipschitz-continuous, and $\lim\limits_{n\to\infty}\frac{1}{N}\|\bf{e}_{t}^{{\phi}\to\gamma}\|^2$ in OAMP exists and is finite.
\end{assumption}

The following lemma was proved in \cite{Rangan2016, Takeuchi2017,Berthier2020} for the IIDG property of OAMP with Lipschitz-continuous NLE $\hat{\phi}_t$, which ensures the correctness of the SE for OAMP.

\begin{lemma} 
[Asymptotic IIDG]\label{The:SE}
Suppose that Assumptions \ref{ASS:Model} and \ref{ASS:phi} hold. For OAMP in \eqref{OAMP_coded}, $\bf{e}_{t}^{\gamma\to{\phi}}$ can be modeled as a sequence of IIDG samples that are independent of $\bf{x}$. The $\gamma_t$ and ${\phi}_t$ in  OAMP can be characterized by the following orthogonal transfer functions: 
\BS\label{Eqn:TF1}\begin{align}
\rho_t &= {\gamma}_{\rm SE}(\vartheta_t)\equiv [\hat{\gamma}_{\rm SE}(\vartheta_t)]^{-1} - \vartheta_t^{-1},\label{Eqn:TF1a}\\
\vartheta_{t+1} & = {\phi}_{\rm SE}(\rho_t)\equiv \big([\hat{\phi}_{\rm SE}(\rho_t)]^{-1}- \rho_t \big)^{-1},\label{Eqn:TF1b}
\end{align}\ES 
with 
\BS\label{Eqn:gamma_mmse}\begin{align}
    \hat{\gamma}_{\rm SE}(\vartheta)&\equiv{\rm mmse} \{\bf{x}|\bf{x}+\sqrt{\vartheta}\bf{z},\Gamma\} \nonumber\\
    &= \tfrac{1}{N}{\mr{tr}}\left\{ [{\rm SNR}\bf{A}^H\bf{A} + \vartheta^{-1}\,\bf{I}]^{-1}\right\},\label{Eqn:gamma_se}\end{align}\begin{align}
     \hat{\phi}_{\rm SE}(\rho)&\equiv {\rm mmse}\{\bf{x}|\sqrt{\rho}\bf{x}+\bf{z},\Phi\},\label{Eqn:phi_se}
\end{align} \ES
where $\bf{z}\sim\mathcal{CN}(\bf{0},\bf{I})$ is independent of $\bf{x}$.
\end{lemma}

The asymptotic IIDG property of OAMP was proved in \cite{Rangan2016, Takeuchi2017} for right-unitarily invariant $\bf{A}$ and separable  $\hat{\phi}_t$. Also, the asymptotic IIDG property of AMP was proved in \cite{Berthier2020} for zero-mean IIDG matrix $\bf{A}$ and non-separable Lipschitz-continuous  $\hat{\phi}_t$.  By combining the results in \cite{Rangan2016, Takeuchi2017} and \cite{Berthier2020}, we obtain Lemma \ref{The:SE}, which establishes the asymptotic IIDG property of OAMP with right-unitarily invariant $\bf{A}$ and non-separable Lipschitz-continuous  $\hat{\phi}_t$.
 
 
 \section{Achievable Rates of OAMP in the Coded LUIS}\label{Sec:rate_OAMP}
This section introduces an equivalent model of state evolution for OAMP, enabling us to derive a variational transfer function (VTF) of OAMP. The VTF serves as a fundamental tool for analyzing the achievable rate of OAMP with FEC decoding, revealing the replica capacity optimality of OAMP. Furthermore, we demonstrate that the conventional receivers exhibit capacity sub-optimal, resulting in a rate loss compared to OAMP.

 \begin{figure}[t]
  \centering
  \includegraphics[width=4cm]{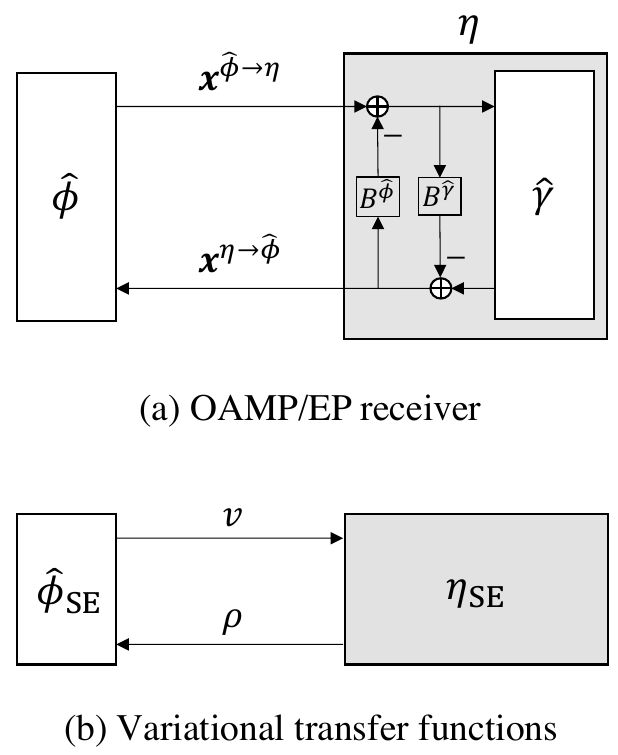}\\ 
  \caption{Graphical illustrations of (a) the OAMP receiver in the form of \eqref{OAMP_simple}, and (b) the VTF given in \eqref{Eqn:SE_equiv}. In (b), the MMSE function $\hat{\phi}_{\rm SE}(\cdot)$ corresponds to the local MMSE estimator $\hat{\phi}_t(\cdot)$ in (a), enabling us to apply the I-MMSE property to OAMP. For simplicity, the memory term and the iterative index $t$ are omitted in the figure.}\label{Fig:SE_OAMP}  
\end{figure}

\subsection{Equivalent Model of State Evolution for OAMP}
 
 The I-MMSE property derived in \cite{Guo2005, Bhattad2007} establishes a connection between achievable rate and MMSE performance. This property will be utilized to examine the replica capacity optimality of OAMP. However, the orthogonal NLE in OAMP is not MMSE-optimal. This poses a challenge in directly applying the I-MMSE property to OAMP with orthogonal estimators in Fig. \ref{Fig:OAMP}. In this subsection, we outline an alternative structure of OAMP, which circumvents this difficulty.

 We rewrite the OAMP in \eqref{Eqn:OAMP} to
 \BS\label{OAMP_simple}\begin{align}
    \bf{x}_{t}^{\eta\to\hat{\phi}} &= \eta_t\big(\bf{x}_{t}^{\hat{\phi} \to\eta}, \bf{x}_{t-1}^{\eta\to\hat{\phi} }\big),\\
   \bf{x}_{t+1}^{\hat{\phi} \to\eta} & = \hat{\phi}_t\big(\bf{x}_{t}^{\eta\to\hat{\phi} }\big), 
\end{align}\ES
where $\eta_t$ includes $\hat{\gamma}_t$ and the GSO operations. As the result, $\eta_t$ contains a memory term $\bf{x}_{t-1}^{\eta\to\hat{\phi} }$. Fig.~\ref{Fig:SE_OAMP}(a) gives a graphical illustration of \eqref{OAMP_simple}.

Let $v_t$ be the MSE of $\bf{x}_{t}^{\hat{\phi} \to\eta}$ and $\rho_t$ the SINR of $\bf{x}_{t}^{\eta\to\hat{\phi} }$:
\BS\begin{align}
  v_t   &\equiv  \tfrac{1}{N} {\rm E}\big\{\| \bf{x}_{t}^{\hat{\phi} \to\eta} -\bf{x}   \|^2\big\},\\
  \rho_t   &\equiv {{N}/{\rm E}\big\{\| \bf{x}_{t}^{\eta\to\hat{\phi} } -\bf{x} \|^2\}}. 
\end{align}\ES
Then, the state evolution for the  equivalent structure of OAMP in \eqref{OAMP_simple} is given by
\BS\label{Eqn:SE_equiv}\begin{align}
\rho_t &=\tilde{\eta}_{\rm SE}(v_{t},\rho_{t-1})= {\gamma}_{\rm SE}\big([v_{t}^{-1} -\rho_{t-1}]^{-1}\big),\label{Eqn:SE_equiva}\\
v_{t+1} & = \hat{\phi}_{\rm SE}(\rho_t),\label{Eqn:SE_equivb}
\end{align}\ES
where ${\gamma}_{\rm SE}(\cdot)$ and $ \hat{\phi}_{\rm SE}(\cdot)$ are given in \eqref{Eqn:TF1a} and \eqref{Eqn:phi_se}, respectively. The additional variable $\rho_{t-1}$ in $\tilde{\eta}_{\rm SE}(\cdot)$ comes from the memory term $\bf{x}_{t-1}^{\eta\to\hat{\phi} }$ contained in $\eta_t(\cdot)$. In \eqref{Eqn:SE_equiv}, the MMSE function $\hat{\phi}^{\cal S}_{\rm SE}(\cdot)$ corresponds to the local MMSE estimator $\hat{\phi}_t(\cdot)$ in \eqref{OAMP_simple}, which enables us to leverage the I-MMSE property to analyze the achievable rate of OAMP.  

\subsection{Variational Transfer Function (VTF) without FEC Coding}\label{Sec:SE} 

For an un-coded LUIS, the NLE $\hat{\phi}_t$ is an MMSE demodulation $\hat{\phi}^{\cal S}_t$. Hence, the OAMP is a detection process given by 
 \BS\label{OAMP_det}\begin{align} 
    \bf{x}_{t}^{\eta\to\hat{\phi}_{\cal S}} &= \eta_t\big(\bf{x}_{t}^{\hat{\phi}_{\cal S} \to\eta}, \bf{x}_{t-1}^{\eta\to\hat{\phi}_{\cal S} }\big), \\ 
   \bf{x}_{t+1}^{\hat{\phi}_{\cal S} \to\eta}  &= \hat{\phi}^{\cal S}_t\big(\bf{x}_{t}^{\eta\to\hat{\phi}_{\cal S}}\big)={\rm E}\{\bf{x}|\bf{x}_{t}^{\eta\to\hat{\phi}_{\cal S}}, {\Phi}_{\cal S}\}, 
 \end{align} \ES
where $\eta_t$ is the same as that in \eqref{OAMP_simple}, and ${\Phi}_{\cal S}$ is the constellation constraint  given in \eqref{Eqn:Phi_S}.  Define 
 \BS\begin{align} 
    \rho_t  &\equiv \big[{\tfrac{1}{N}{\rm E}\big\{\|  \bf{x}^{\hat{\phi}_{\cal S}\to\eta}_t -\bf{x} \|^2\}}\big]^{-1},\\
   v_t  &\equiv \tfrac{1}{N}{\rm E}\big\{\| \bf{x}^{\eta \to\hat{\phi}_{\cal S}}_t -\bf{x} \|^2\big\}.
\end{align} \ES
For IID un-coded $\bf{x}$, Assumptions \ref{ASS:Model} and \ref{ASS:phi} strictly hold \cite[Lemma 2]{Takeuchi2017}. Therefore, following Lemma \ref{The:SE}, the state evolution for the OAMP in \eqref{OAMP_simple} is given by
 \BS\begin{align} 
\rho_t  &=\tilde{\eta}_{\rm SE}(v_{t},\rho_{t-1}),\\
v_{t+1} &= \hat{\phi}_{\rm SE}^{\cal S}(\rho_t) = {\rm mmse}\{\bf{x}|\sqrt{\rho_t}\bf{x}+\bf{z},\Phi_{\cal S}\},
\end{align} \ES
where $\tilde{\eta}_{\rm SE}$ is given in \eqref{Eqn:SE_equiva}, ${\Phi}_{\cal S}$ is given in \eqref{Eqn:Phi_S}, and $\bf{z}\sim\mathcal{CN}(\bf{0},\bf{I})$. For Gaussian signaling \cite{Guo2005random},  
\BE
    \hat{\phi}_{\rm SE}^{\cal S}(\rho_t)=\hat{\phi}_{\rm SE}^{\rm Gau}(\rho_t)= {1}/(1+\rho_t).
\EE

\emph{New Challenge:} Despite the advantages offered by the new structure for analyzing the achievable rate of OAMP, it also introduces a new challenge. The transfer function $\tilde{\eta}_{\rm SE}(\cdot)$ now becomes a dual-input-single-output (DISO) function, incorporating an additional variable $\rho_{t-1}$ from the previous iteration. This poses difficulties in conducting the achievable rate analysis of OAMP. As a consequence, the I-MMSE property and the results in \cite{LeiTIT} for AMP, which are specifically applicable to SISO transfer curve $\tilde{\eta}_{\rm SE}(\cdot)$, become infeasible for OAMP.

To solve the new challenge, we relax the DISO function $\tilde{\eta}_{\rm SE}(\cdot)$ to a SISO function ${\eta}_{\rm SE}(\cdot)$ by replacing $\rho_{t-1}$ with $\rho_t$. Then we obtain the following VTF for the OAMP.
\BS\label{Eqn:TF}\begin{align}
\rho  &= {\eta}_{\rm SE}(v)\equiv v^{-1}-[\hat{\gamma}_{\rm SE}^{-1}(v)]^{-1},\\ 
v   &= \hat{\phi}_{\rm SE}^{\cal S}(\rho), 
\end{align}\ES
where $\hat{\gamma}_{\rm SE}^{-1}(\cdot)$ is the inverse of $\hat{\gamma}_{\rm SE}(\cdot)$. Fig. \ref{Fig:TF_chart} provides a graphical illustration of the VTF in \eqref{Eqn:TF} for OAMP detection. Based on the VTF, we can derive several important area properties of LUIS.   

 \emph{Notes}: The state evolution in \eqref{Eqn:SE_equiv} is equivalent to that in \eqref{Eqn:TF1} by substitution $\vartheta_t=[v_{t}^{-1} -\rho_{t-1}]^{-1}$. Therefore,  the MSE performance of OAMP at each iteration can be characterized using either \eqref{Eqn:TF1} or \eqref{Eqn:SE_equiv}. The VTF in \eqref{Eqn:TF}, however, are not equivalent to those in \eqref{Eqn:TF1} and \eqref{Eqn:SE_equiv}. Hence, using \eqref{Eqn:TF} to characterize the MSE performance of OAMP at each iteration is no longer applicable. Nevertheless, as will be shown in Section \ref{Sec:rate_OAMP}, VTF plays a crucial role in the achievable analysis of OAMP.

 \subsection{Variational Transfer Function (VTF) with FEC Coding}\label{Sec:VT_cod}

For a LUIS with FEC coding, we rewrite the problem as
\BS\begin{align}
 {\rm Linear\; constraint}\;\; \Gamma: \quad& \bf{y}=\bf{Ax}+\bf{n},\\
{\rm Code\; constraint}\;\;  \Phi_{\cal C}: \quad& \bf{x}\in  \bf{\mathcal{C}}, \quad x_i\sim P_X(x),\; \forall i,\label{Eqn:code}
\end{align}\ES
where $\bf{\mathcal{C}}$ is a codebook. We focus on a joint OAMP and \emph{a-posteriori} probability (APP) decoding for a coded LUIS.
\BS\label{OAMP_coded} \begin{align}
    \bf{x}_{t}^{\eta\to\hat{\phi}_{\cal C}} &= \eta_t\big(\bf{x}_{t}^{\hat{\phi}_{\cal C}\to\eta}, \bf{x}_{t-1}^{\eta\to\hat{\phi}_{\cal C}}\big),\\
   \bf{x}_{t+1}^{\hat{\phi}_{\cal C}\to\eta} & = \hat{\phi}_{t}^{\cal C}\big(\bf{x}_{t}^{\eta\to\hat{\phi}_{\cal C}}\big) ={\rm E}\{\bf{x}|\bf{x}^{\gamma\to \phi}_t,\Phi_{\mathcal{C}}\}, 
\end{align}\ES
where $\eta_t$ is the same as that in \eqref{OAMP_simple}.  In contrast to the un-coded LIUS, where a symbol-wise demodulator $\hat{\phi}_{t}^{\cal S}$ is used, the coded LIUS employs an APP decoder $\hat{\phi}_{t}^{\cal C}$.

In this paper, we assume that different bits in the FEC code are asymptotically pairwise independent, which implies $\lim\limits_{n\to\infty}\frac{1}{N}\|{\bm{x}}\|^2\overset{\rm a.s.}{=}\frac{1}{N}{\rm E}\{\|{\bm{x}}\|^2\}= 1$. Hence, Assumption \ref{ASS:Model} is satisfied. This assumption is widely used for the analysis of Turbo/LDPC codes with asymptotically long random interleaving \cite{Urbanke2008}. Also, we consider a Lipschitz-continuous APP decoder $\phi_t^{\cal C}(\cdot)$ (i.e., MMSE de-noisier), whose output can be modeled by an MMSE model, i.e., $\bm{x}_t^{\hat{\phi}_{\cal C}\to\eta} = \bm{x} - \bm{e}_t^{\hat{\phi}_{\cal C}\to\eta}$, where $\frac{1}{N}\langle \bm{x}_t^{\hat{\phi}_{\cal C}\to\eta}, \bm{e}_t^{\hat{\phi}_{\cal C}\to\eta} \rangle \overset{\rm a.s.}{\to}0$. Then, we have $\frac{1}{N}{||\bm{e}_t^{\hat{\phi}_{\cal C}\to\eta}||^2} \overset{\rm a.s.}{=} \frac{1}{N}||\bm{x}||^2 - \frac{1}{N}||\bm{{x}}_t^{\hat{\phi}_{\cal C}\to\eta}||^2 <1$ since  $\frac{1}{N}\|{\bm{x}}\|^2\overset{\rm a.s.}{\to}1$ and $\frac{1}{N}\langle \bm{x}_t^{\hat{\phi}_{\cal C}\to\eta}, \bm{e}_t^{\hat{\phi}_{\cal C}\to\eta} \rangle \overset{\rm a.s.}{\to}0$. Hence, $\lim\limits_{n\to\infty}\frac{1}{N}\|\bm{e}_{t}^{{\phi}\to\gamma}\|^2$ exists and is finite. This implies that Assumption \ref{ASS:phi} is satisfied. By satisfying these assumptions, all the preconditions of Lemma \ref{The:SE} are met. Consequently, the IIDG property, as stated in Lemma \ref{The:SE}, holds for the joint OAMP and APP decoding in \eqref{OAMP_coded}. Specifically, for LDPC codes, the Lipschitz continuity of the APP decoder $\hat{\phi}_{t}^{\cal C}$ was proved in \cite[Appendix B]{Ebert2023} under a widely used sub-girth condition. This condition ensures that fewer message-passing iterations are performed on the factor graph of the LDPC code than the shortest cycle of the same graph per OAMP iteration. Moreover, the IIDG properties of OAMP with convolutional decoders and LDPC decoders have also been verified through simulation results in \cite{MaTWC} (e.g., Fig. 4 and Fig. 7 in \cite{MaTWC}).

Define \vspace{-2mm}
\BS\label{Eqn:errors_codes} \begin{align}
    \rho_t  &\equiv \big[{\tfrac{1}{N}{\rm E}\big\{\|  \bf{x}^{\hat{\phi}_{\cal C}\to\eta}_t -\bf{x} \|^2\}}\big]^{-1},\\
   v_t  &\equiv \tfrac{1}{N}{\rm E}\big\{\| \bf{x}^{\eta \to\hat{\phi}_{\cal C}}_t -\bf{x} \|^2\big\}.  
\end{align}\ES
Then, following Lemma \ref{The:SE}, the state evolution for the  equivalent structure of OAMP in \eqref{OAMP_coded} is given by 
\BS\label{Eqn:SE_equiv_C}\begin{align}
\rho_t &=\tilde{\eta}_{\rm SE}(v_{t},\rho_{t-1})= {\gamma}_{\rm SE}\big([v_{t}^{-1} -\rho_{t-1}]^{-1}\big),\\
v_{t+1} & = \hat{\phi}_{\rm SE}^{\cal C}(\rho_t)={\rm mmse}\{\bf{x}|\sqrt{\rho}\bf{x}+\bf{z},\Phi_{\cal C}\}.\label{Eqn:SE_equiv_Cb}
\end{align}\ES
where ${\gamma}_{\rm SE}(\cdot)$ is given in \eqref{Eqn:TF1a}.

Considering that $\hat{\phi}_{\text{SE}}^{\mathcal{C}}(\cdot)$ is an MMSE function, the I-MMSE property presented in \cite{Guo2005, Bhattad2007} can be utilized to analyze the achievable rate of OAMP. However, the presence of the memory term $\rho_{t-1}$ in $\tilde{\eta}_{\text{SE}}(\cdot)$ makes it challenging to perform the achievable rate analysis. To facilitate the analysis, similar to the un-coded case in Subsection \ref{Sec:SE}, we relax $\tilde{\eta}_{\rm SE}(\cdot)$ to a SISO function by replacing $\rho_{t-1}$ with $\rho_t$. This relaxation results in the following VTF of the OAMP in \eqref{OAMP_coded}:  
 \BS\label{Eqn:TF_C}\begin{align}
  \rho & = {\eta}_{\rm SE}(v)\equiv v^{-1}-[\hat{\gamma}_{\rm SE}^{-1}(v)]^{-1},\\  
 v &= \hat{\phi}_{\rm SE}^{\cal C}(\rho).
 \end{align}\ES

\subsection{Area Property} 
In the un-coded case, OAMP is not error-free and converges to a non-zero fixed point $(\rho^*,v^*)$  (see Fig. \ref{Fig:TF_chart}).  In the coded case,  error-free recovery is possible if $\hat{\phi}_t^{\cal C}(\cdot)$ is properly designed. Following Lemma \ref{The:SE}, the proposition below specifies the sufficient and necessary condition for an error-free OAMP.
\begin{proposition}\label{Pro:ZF_EXIT}
Suppose that Assumptions  \ref{ASS:Model}-\ref{ASS:phi} hold. The OAMP in \eqref{OAMP_coded} achieves error-free recovery if and only if there exists a tunnel between the orthogonal transfer functions ${\phi}_{\rm SE}^{\cal C}(\rho)$ and ${\gamma}_{\rm SE}^{-1}(\rho)$ that converges to zero variance, and there is no fixed point between ${\phi}_{\rm SE}^{\cal C}(\rho)$ and ${\gamma}_{\rm SE}^{-1}(\rho)$. That is,
\BE\label{Eqn:upper_bound_EXIT}
      {\phi}_{\rm SE}^{\cal C}(\rho) < {\gamma}_{\rm SE}^{-1}(\rho),\;\;\; 0 \leq {\rho } < {\gamma}_{\rm SE}(0), 
\EE 
where ${\gamma}_{\rm SE}^{-1}$ is the inverse of $\gamma_{\rm SE}$ is given in \eqref{Eqn:TF1a},  ${\gamma}_{\rm SE}(0) ={\rm SNR}$,  and
\BE
    {\phi}_{\rm SE}^{\cal C}(\rho) = \big([\hat{\phi}^{\cal C}_{\rm SE}(\rho_t)]^{-1}- \rho_t \big)^{-1}.
\EE
\end{proposition}

By leveraging the code-rate-MMSE lemma in \cite{Bhattad2007} and the sufficient and necessary error-free condition in Proposition \ref{Pro:ZF_EXIT}, we can derive the achievable rate of OAMP as follows.

\begin{lemma}\label{Lem:Rate_orth}
Suppose that Assumptions  \ref{ASS:Model}-\ref{ASS:phi} hold. The achievable rate of an OAMP receiver is given by 
\BS\label{Eqn:R_OAMP}\begin{align}
 & R_{\rm OAMP }({\rm SNR}) = \int_{0}^{\infty} \hat{\phi}_{\rm SE}^{\cal C}(\rho) d\rho,  \\
  {\rm s.t.}\quad &{\phi}_{\rm SE}^{\cal C}(\rho) < {\gamma}_{\rm SE}^{-1}(\rho),\;\;\; 0 \leq {\rho } < {\gamma}_{\rm SE}(0). \label{Eqn:R_OAMP_cond} 
\end{align}\ES
That is, the OAMP receiver achieves error-free recovery if $R_{\cal C}\leq R_{\rm OAMP}$.
\end{lemma}

Fig. \ref{Fig:track} provides a graphic illustration of Lemma \ref{Lem:Rate_orth}. In the next subsections, we will rephrase the error-free constraint in \eqref{Eqn:R_OAMP_cond} using VTF  (see Lemmas \ref{Lem:ZF_equal}, \ref{Lem:ZF_APP} and \ref{Lem:Rate_VTF}).

\subsection{VTF Properties}\label{Sec:area_LUIS}

Proposition \ref{Pro:Mon_decr} establishes the monotonicity of $ {\eta}_{\rm SE}$.
\begin{proposition}\label{Pro:Mon_decr}
   ${\eta}_{\rm SE}(v)$ is a strictly decreasing function in $v\ge 0$.
\end{proposition}
\begin{IEEEproof}
    Following \eqref{Eqn:TF1} and \eqref{Eqn:TF}, we rewrite ${\eta}_{\rm SE}(v)$  to 
     \BE
        {\eta}_{\rm SE}(v) = {\gamma}_{\rm SE}\big( \hat{\gamma}_{\rm SE}^{-1}(v) \big).  
     \EE
     It is demonstrated in \cite{Ma2016} that ${\gamma}_{\rm SE}$ is a strictly  decreasing function and $ \hat{\gamma}_{\rm SE}$ is a strictly increasing function, implying that $ \hat{\gamma}^{-1}_{\rm SE}$ is a strictly increasing function. Therefore, ${\eta}_{\rm SE}$, being the composition of a strictly decreasing function and a strictly increasing function, is a strictly decreasing function.
\end{IEEEproof}

\begin{lemma}\label{Lem:same_FP} 
The fixed-point equation  $\hat{\phi}_{\rm SE}^{\cal S}(\rho)= {\eta}^{-1}_{\rm SE}(\rho)$ for the VTF in \eqref{Eqn:TF} is equivalent to the fixed-point equation ${\phi}_{\rm SE}^{\cal S}(\rho)= {\gamma}^{-1}_{\rm SE}(\rho)$ for the orthogonal transfer functions in \eqref{Eqn:TF1}. 
\end{lemma}
\begin{IEEEproof} 
    See Appendix \ref{APP:same_FP}.
\end{IEEEproof}

 The following is crucial for the achievable rate of OAMP.
\begin{lemma}\label{Lem:ZF_equal}
 For $\rho \ge 0$, ${\phi}_{\rm SE}^{\cal C}(\rho)<{\gamma}^{-1}_{\rm SE}(\rho)$ holds if and only if $ \hat{\phi}_{\rm SE}^{\cal C}(\rho)<{\eta}^{-1}_{\rm SE}(\rho)$ holds.  
\end{lemma} 
\begin{IEEEproof} 
   See Appendix \ref{APP:ZF_equal}.
 \end{IEEEproof}
 Following Proposition \ref{Pro:ZF_EXIT} and Lemma \ref{Lem:ZF_equal}, we can figure out the necessary and sufficient condition of the VTF for an error-free OAMP. 
 \begin{lemma}\label{Lem:ZF_APP}
Suppose that Assumptions  \ref{ASS:Model}-\ref{ASS:phi} hold. As shown in Fig. \ref{Fig:track}, OAMP achieves error-free recovery\footnote{In fact, we can never do an error-free recovery with finite-length coding. In practice, we can change the error condition to 
\BE
    \hat{\phi}_{\rm SE}^{\cal C}(\rho) <  {\eta}_{\rm SE}^{-1}({\rho}), \;\;\; 0 \leq {\rho } \leq {\eta}_{\rm SE}(\epsilon), \nonumber
\EE
where $\epsilon$ is the target error of the OAMP receiver. In the simulations of this paper, we set $\epsilon$ in $10^{-4}\sim10^{-5}$.} if and only if  
\BE\label{Eqn:upper_bound}
     \hat{\phi}_{\rm SE}^{\cal C}(\rho) < \hat{\phi}_{\rm SE}^*(\rho)\equiv \min\{{\eta}_{\rm SE}^{-1}({\rho}),\; \hat{\phi}_{\rm SE}^{\cal S}(\rho)\},
\EE 
for $ 0 \leq {\rho } < {\rm SNR}$, where $ \hat{\phi}_{\rm SE}^{\cal C}(\rho) <  \hat{\phi}_{\rm SE}^{\cal S}(\rho)$ follows that the decoding MSE should be lower than that of the detector, and $\hat{\phi}_{\rm SE}^{\cal C}(\rho) <  {\eta}_{\rm SE}^{-1}({\rho})$ follows the error-free condition (see Proposition \ref{Pro:ZF_EXIT} and Lemma \ref{Lem:ZF_equal}).
\end{lemma} 

Lemma \ref{Lem:ZF_APP} plays a crucial role in the analysis of the achievable rate and in proving the replica capacity optimality of OAMP in the subsequent subsection.

\subsection{Achievable Rate of Coded OAMP System}\label{Sec:area_AMP} 

\begin{assumption}\label{Pro:SCP}
There is exactly one fixed point for $\hat{\phi}_{\rm SE}^{\cal S}(\rho) = {\eta}^{-1}_{\rm SE}(\rho)$ in $\rho>0$.
\end{assumption}

The convergence of the SE of OAMP was demonstrated in \cite{Takeuchi2021OAMP, SS_MAMPISIT,Lei_SSMAMP}. Suppose that OAMP converges to a unique fixed point $(\rho^*, v^*)$ with $v^*=\hat{\phi}_{\rm SE}^{\cal S} (\rho^*)$. The following theorem establishes the area property of a LUIS.

\begin{theorem}[Capacity-Area Theorem]\label{The:area_LUIS}
  Suppose that Assumptions  \ref{ASS:Model} and \ref{Pro:SCP} hold. The replica constrained capacity of LUIS with a fixed $p_X(x)$ is determined by the area $A_{\rm ADGO}$ (see Fig. \ref{Fig:TF_chart}) covered by ${\eta}_{\rm SE}^{-1}$ and $\hat{\phi}_{\rm SE}^{\cal S}$, i.e.,  
\BE\label{Eqn:dis_cap}
     C_{\rm Rep}({\rm SNR}) = A_{\rm ADGO}, 
\EE
 where 
\BS \label{Eqn:A_C} \begin{align}
      &A_{\rm ADGO}  =  \int_{0}^{\rho^*}  {\hat{\phi}_{\rm SE}^{\cal S}(\rho) \,d\rho} + \int_{\rho^*}^{{\rm SNR}}  {{\eta}_{\rm SE}^{-1}(\rho) \,d\rho}  \label{Eqn:A_C1} \\
     &= \int_{0}^{\rho^*}\!\!\!  {\hat{\phi}_{\rm SE}^{\cal S}(\rho) \,d\rho} +\log  v^* \! +  { \tfrac{1}{N} \log \det\big( \bf{B}(\rho^*, v^*) \big)},  \label{Eqn:A_C2}
\end{align} 
\ES
and $\bf{B}(\rho^*, v^*) = \big([v^*]^{-1} - \rho^*\big)\bf{I} + {\rm SNR}\bf{A}^H\bf{A}$.  
\end{theorem}  

\begin{IEEEproof} 
   See Appendix \ref{APP:area_express}.
\end{IEEEproof}

Following Lemma \ref{Lem:ZF_APP}, the achievable rate of the OAMP in Lemma \ref{Lem:Rate_orth} can be rewritten into a VTF form as follows.
  
\begin{lemma}\label{Lem:Rate_VTF}
Suppose that Assumptions  \ref{ASS:Model}-\ref{ASS:phi} hold. The achievable rate of an OAMP receiver can be expressed as
\BS\begin{align}
 & R_{\rm OAMP }({\rm SNR}) = \int_{0}^{\infty} \hat{\phi}_{\rm SE}^{\cal C}(\rho) d\rho,  \\
  {\rm s.t.}\quad & \hat{\phi}_{\rm SE}^{\cal C}(\rho) <  \hat{\phi}_{\rm SE}^*(\rho),\;\;\; 0 \leq {\rho } < {\gamma}_{\rm SE}(0).  
\end{align}\ES 
\end{lemma}  

The theorem below follows Theorem \ref{The:area_LUIS} and Lemma \ref{Lem:Rate_VTF}.
 
\begin{theorem}[Replica Capacity Optimality]\label{The:cap_opt}
 Suppose that Assumptions  \ref{ASS:Model}-\ref{Pro:SCP} hold. The achievable rate of OAMP achieves the replica constrained capacity of LUIS, i.e., 
\BE\label{Eqn:R_to_C}
R_{\rm OAMP}({\rm SNR})\to C_{\rm Rep}({\rm SNR}), 
\EE 
if $\hat{\phi}_{\rm SE}^{\cal C}(\rho) < \hat{\phi}_{\rm SE}^*(\rho)$ and $\hat{\phi}_{\rm SE}^{\cal C}(\rho) \to \hat{\phi}_{\rm SE}^*(\rho)$ in $[0,  {\rm SNR}]$.  
\end{theorem}

 For IIDG matrices \cite{Reeves_TIT2019, Barbier2017arxiv} and certain sub-class right-unitarily-invariant matrices \cite{Barbier2018b, Fan2022},  $C_{\rm Rep}({\rm SNR})$ denotes the true constrained capacity of LUIS. In this case, OAMP is rigorously capacity optimal. In addition, Theorem \ref{The:cap_opt} is developed under the matching constraint $ \hat{\phi}_{\rm SE}^{\cal C}(\rho) \to \hat{\phi}_{\rm SE}^*(\rho)$. The existence of a curve-matching code is proved in Appendix C-B \cite{LeiTIT} for Gaussian signaling. However, for non-Gaussian signaling, the existence of a curve-matching code still remains a conjecture. Some numerical results will be provided in Section \ref{Sec:SIM} to empirically verify the conjecture for QPSK modulations.


 \subsection{Special Instance: Gaussian Signaling}
 In this subsection, we focus on a special case that $\bf{x}$ is Gaussian. Note that the unique fixed point assumption (see Assumption \ref{Pro:SCP}) cannot always be guaranteed for general signaling. However, we demonstrate that with Gaussian signaling, Assumption \ref{Pro:SCP} holds asymptotically. Furthermore, the existence of a curve-matched code can be verified via superposition-coded modulation (SCM). Additionally, the constrained capacity is reduced to the well-known Gaussian capacity. 

\begin{lemma}[Unique Fixed Point]\label{Lem:uni_point}
 For a LUIS with $\bf{x}\sim \mathcal{CN}(\bf{0},\bf{I})$, Assumption \ref{Pro:SCP} holds asymptotically. More specifically, the positive solution of $\hat{\phi}_{\rm SE}^{\cal S}(\rho) =  {\eta}_{\rm SE}^{-1}({\rho})$ is unique and can be explicitly expressed as 
    \BE\label{Eqn:Gau_rho}
      \rho^{*}_{\mr{Gau}}= \big[\hat{\gamma}_{\rm SE}(1)\big]^{-1} -1. 
    \EE   
\end{lemma}

\begin{IEEEproof}
For $\bf{x}\sim \mathcal{CN}(\bf{0},\bf{I})$, we have $\hat{\phi}_{\rm SE}^{\cal S}(\rho)=\hat{\phi}_{\rm SE}^{\rm Gau}(\rho)= {1}/(1+\rho)$ \cite{Guo2005random}. Substituting it and \eqref{Eqn:TF} into $\hat{\phi}_{\rm SE}^{\rm Gau}(\rho) =  {\eta}_{\rm SE}^{-1}({\rho})$, we obtain \eqref{Eqn:Gau_rho}. Since $\hat{\gamma}_{\rm SE}(1)={\rm mmse}\{\bf{x}|\bf{x}+\bf{z},\Gamma\}$ is monotonous and less than 1, $\rho^*_{\rm Gau}$ is positive and unique.
\end{IEEEproof}

The proof of the following lemma is omitted as it is the same as that in \cite[Appendix C-B]{LeiTIT}.
  
\begin{lemma}[Existence of Curve-Matched Code]\label{Lem:cod_exist}
For $\bf{x}\sim \mathcal{CN}(\bf{0},\bf{I})$, there exists a superposition coded modulation  whose transfer function $\hat{\phi}_{\rm SE}^{\cal C}(\rho) \to \hat{\phi}_{\rm SE}^*(\rho) $ in $[0, {\rm SNR}]$.
\end{lemma}

Following Lemma \ref{Lem:uni_point} and Lemma \ref{Lem:cod_exist}, we have the following theorem for Gaussian signaling.
 
\begin{theorem}\label{The:cap_opt_Gau}
 The Gaussian capacity of LUIS is given by $A_{\rm ACGO}$, i.e.,  
\BE\label{Eqn:cap_Gau}
  \!\!\!  A_{\rm ACGO}\!=\!C_{\mr{Gau}}({\rm SNR})\!=\! \tfrac{1}{N}\log\det\big(\bf{I}\! + \! {\rm SNR}\bf{A}^H\!\bf{A}\big). 
\EE  
Suppose that Assumptions  \ref{ASS:Model}-\ref{ASS:phi} hold. When  $\hat{\phi}^{\mathcal{C}}_t$ approaches $\hat{\phi}_{\rm SE}^{\rm Gau}(\rho) $, the achievable rate of OAMP approaches the Gaussian capacity, i.e., $R_{\rm OAMP}({\rm SNR})\to C_{\mr{Gau}}({\rm SNR})$.
\end{theorem}  

\begin{IEEEproof}
     For $\bf{x}\sim \mathcal{CN}(\bf{0},\bf{I})$, we have $\hat{\phi}_{\rm SE}^{\cal S}(\rho)=\hat{\phi}_{\rm SE}^{\rm Gau}(\rho)={1}/(1+\rho)$. Thus, $\int_{0}^{\rho^*} \hat{\phi}_{\rm SE}^{\cal S}(\rho) \,d\rho +\log \hat{\phi}_{\rm SE}^{\cal S}(\rho^*) =0$. Then, following Theorem \ref{The:cap_opt}, we have 
 \BE
R_{\mr{OAMP}}({\rm SNR}) \!=\!  A_{\rm ADGO}  
\! = \!  \tfrac{1}{N} \log \det\big(\bf{I} + {\rm SNR}\bf{A}^H\bf{A} \big), 
\EE 
which is the same as the Gaussian capacity in \eqref{Eqn:Gau_C}.
\end{IEEEproof}

\subsection{Comparisons with Conventional Methods} \label{Sec:Comparisons}  
  
\emph{1) Comparing with Conventional Turbo:} Turbo is extrinsic and requires independent input-output errors for each local processor. In contrast, OAMP only necessitates orthogonal input-output errors, which are generally less stringent than the independent requirement. It was proved in \cite{MaTWC} that the MSE of OAMP is lower than Turbo (whose achievable rate was given in \cite{Yuan2014}). That is, ${\eta}^{-1}_{\mr{SE-OAMP}}\geq {\eta}^{-1}_{\mr{SE-Turbo}}$. Consequently, based on the I-MMSE lemma, the achievable rate of OAMP is not lower than Turbo.   

\emph{2) Comparing with Cascading OAMP:}
A cascading OAMP (CAS-OAMP) receiver, as defined in \cite{Guo2005random, Tanaka2002}, operates by first using OAMP for detection and then utilizing its result for decoding, without any iteration between the two stages. The achievable rate of CAS-OAMP is indicated by Area $A_{\rm ADEO}$ in Fig. \ref{Fig:TF_chart}: $A_{\rm ADEO} = R_{\rm CAS}({\rm SNR}) = \int_{0}^{\rho^*} \hat{\phi}_{\rm SE}^{\cal S}(\rho) d\rho$. For Gaussian signaling, $R_{\mr{CAS}}   = \log(1+\rho^{*}_{\mr{Gau}})$. Comparing with the $R_{\rm OAMP}$ in Theorem \ref{The:cap_opt} (see \eqref{Eqn:dis_cap} and \eqref{Eqn:R_to_C}), the rate loss of CAS-OAMP can be quantified by the area $A_{\rm DGE}$ in Fig. \ref{Fig:TF_chart}: $\Delta R_{\mr{CAS}}=  A_{\rm DGE} =  \log  v^* + \!{ \tfrac{1}{N} \log \det\big( \bf{B}(\rho^*) \big)}$.

\subsection{Summary}\label{Sec:areas}

 The main findings in the paper can be summarized by the following area properties of LUIS, as depicted in Fig.~\ref{Fig:TF_chart}. 
\begin{enumerate}   
\item Area $A_{\rm ADGO}$ equals the replica constrained capacity $C_{\rm Rep}$ of a LUIS. See Theorem \ref{The:area_LUIS}.

\item Area $A_{\rm ACGO}$ equals the Gaussian capacity $C_{\rm \mr{Gau}}$ of a LUIS. See Theorem \ref{The:cap_opt_Gau}.   
\BE
    A_{\rm ACGO} = C_{\rm Gau} ({\rm SNR}). 
\EE
As a result, area $A_{\rm ACD}$ represents the  rate gain of Gaussian signaling.  
 
\item Area $A_{\rm ADEO}$ equals the achievable rate of a cascading receiver with OAMP detection  and decoding. That is, 
\BE
    A_{\rm ADEO} = R_{\rm CAS-OAMP}({\rm SNR})  = C_{\rm SISO}(\rho^*). 
\EE
Hence, area $A_{\rm DGE}$ represents the rate loss of a cascading receiver, i.e.,  
  \BE
        A_{\mr{DGE}}  = C_{\rm Rep}({\rm SNR}) -  R_{\rm CAS-OAMP}({\rm SNR}).
\EE 

\item Area $A_{\rm AIJO}$ equals the achievable rate of a cascading receiver with (non-iterative) LMMSE  detection and decoding. That is, 
\BE
    A_{\rm AIJO} = R_{\rm CAS-LMMSE}({\rm SNR})  = C_{\rm SISO}(\rho_0). 
\EE
Hence, area $A_{\rm IDEJ}$ represents the rate loss of the non-iterative LMMSE receiver, i.e., 
\BE
   \!\!\! A_{\rm IDEJ} = R_{\rm CAS-OAMP}({\rm SNR}) - R_{\rm CAS-LMMSE}({\rm SNR}). 
\EE

\item Area $A_{\rm AFGO}$ equals the constrained capacity of a SISO channel, i.e., $A_{\rm AFGO} = C_{\rm SISO}({\rm SNR}) = \int_{0}^{{\rm SNR}} \hat{\phi}_{\rm SE}^{\cal S}(\rho) d\rho$. Hence, area $A_{\rm DFG}$ represents the capacity gap of parallel SISO channels and LUIS, i.e., the rate loss caused by the cross-symbol interference in $\bf{Ax}$: 
\BE
   A_{\rm DFG}  = C_{\rm SISO}({\rm SNR}) - C_{\rm Rep}({\rm SNR}).
\EE

\item Area $A_{\rm AHO}$ equals the constellation entropy: $A_{\rm AHO} = \log|\mathcal{S}|$, i.e., the rate of the noiseless case. Hence, area $A_{\rm FHG}$ represents the rate loss caused by  the channel noise $\bf{n}$, i.e.,  
 \BE
   A_{\rm FHG}  \!=\! A_{\rm AHO}  - A_{\rm AFGO}  \! =\! \log|\mathcal{S}| - C_{\rm SISO}({\rm SNR}).
\EE  

\item  The replica constrained capacity of LUIS can be rewritten to $C_{\rm Rep}=A_{\rm BDGO}+ A_{\rm ADEO} - A_{\rm BDEO}$, where $ A_{\rm BDGO} = A_{\rm BDEO} + A_{\rm DGE}   = \rho^*v^* +  \log  v^* + \tfrac{1}{N} \log \det\big( \bf{B}(\rho^*) \big)$.  Since $A_{\rm BDEO}=\rho^*v^*$ and $A_{\rm ADEO}= C_{\rm SISO}(\rho^*)= I(x;\sqrt{\rho^*}x+z)$, following \eqref{Eqn:dis_cap_ref}, we have
\BE
    A_{\rm BDGO}=\int_0^{{\rm SNR}\,v^*} \mathcal{R}_{\bf{A}^H\bf{A}}(-z)dz,
\EE
where $\mathcal{R}_{\bf{A}^H\bf{A}}(\cdot)$ is the \emph{R-transform} given in \eqref{Eqn:R_transform}.
\end{enumerate}

 \begin{figure*}[t] \vspace{-0.5cm}
  \centering
  \includegraphics[width=15cm, height=4.3cm]{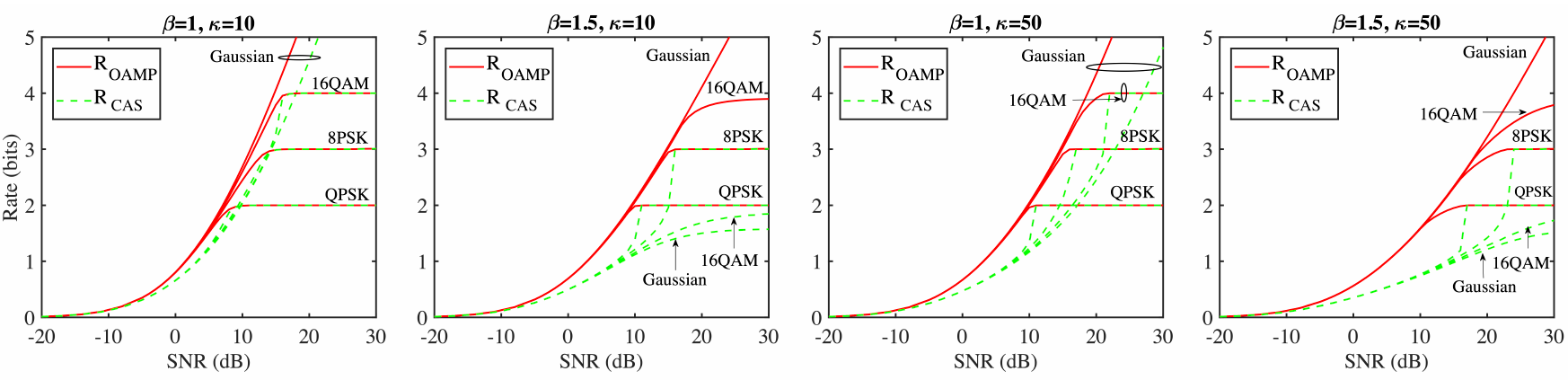}\\  
  \caption{Achievable rate comparison between CAS-OAMP and OAMP in LUIS with $\beta=N/M=\{1, 1.5\}$ and $\kappa=\{10, 50\}$ (condition numbers). QPSK, 8PSK, 16QAM, and Gaussian modulations are considered. Matrix $\bf{A}$ is generated in the same way as that in Section \ref{Sec:SIM}.}\label{Fig:Rate_CDD}  
\end{figure*}

\section{Discussions}

\begin{figure*}[t] 
  \centering
  \includegraphics[width=14cm]{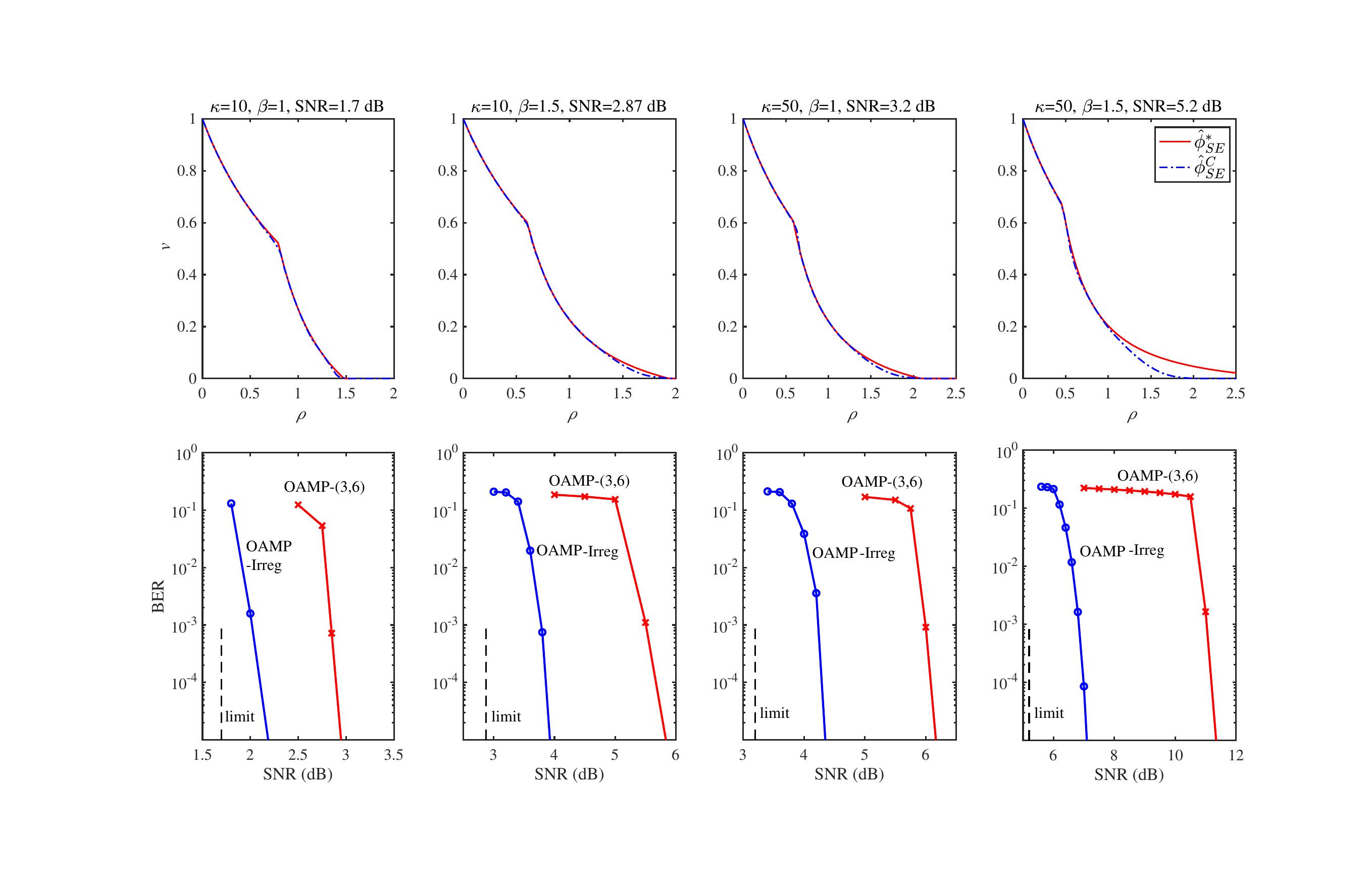}\\ 
  \caption{Curve matching and BER performances of OAMP, where ``$\hat{\phi}_{\rm{SE}}^*$''  is target curve, ``$\hat{\phi}_{\rm{SE}}^{\mathcal{C}}$'' the optimized curve, ``OAMP-Irreg'' the OAMP with optimized irregular LDPC codes, ``OAMP-(3, 6)'' the OAMP with regular (3, 6) LDPC codes. QPSK modulation, codeword length = $10^5$, code rate $\approx$ 0.5, and iterations = $250$. See Table \ref{Opt_degree1} for the details.}\label{Fig:BER_OAMP}  
\end{figure*}

\emph{1) MMSE Optimality vs Capacity Optimality}: Previous works, such as \cite{Tulino2013, Kabashima2006, Ma2016}, have established the MMSE-optimality of OAMP in LUIS using state evolution. In this paper, we establish the capacity-optimality of OAMP in LUIS through appropriate code design.  It is important to note that capacity optimality and MMSE optimality are distinct concepts. The capacity optimality of OAMP is, at the very least, not a direct result of the MMSE optimality. In addition to MMSE optimality, the capacity optimality of OAMP relies on the optimal coding principle found in this paper. For instance, the MMSE-optimal OAMP with regular $(3, 6)$ LDPC codes falls significantly short of the LUIS capacity limit (see Fig. \ref{Fig:BER_OAMP}), and the MMSE-optimal OAMP with ideal SISO codes (e.g., irregular LDPC codes) exhibits considerable rate loss (see \ref{Sec:Comparisons} and Fig. \ref{Fig:Rate_CDD}).

\emph{2) Differences from the OAMP for Coded Linear Systems in \cite{MaTWC}}:  The concept of using the OAMP receiver for coded linear systems was first introduced in \cite{MaTWC}. It was rigorously demonstrated that the MSE performance of OAMP is not worse than the state-of-the-art Turbo-LMMSE (Wang and Poor) method for coded linear systems \cite{MaTWC}. However, \cite{MaTWC} did not investigate the achievable rate, the optimal coding principle, or the replica capacity optimality of OAMP. To the best of our knowledge, this is the first study to analyze the achievable rate performance, develop the optimal coding principle, and demonstrate the replica capacity optimality of OAMP for coded linear systems.  Notably, we show that OAMP with optimized codes significantly outperforms OAMP with unoptimized codes, such as the SISO regular/irregular LDPC codes (see Fig. \ref{Fig:Rate_CDD} and Fig. \ref{Fig:BER_OAMP}).

\emph{3) Differences from the Capacity Optimality of AMP in \cite{LeiTIT}}: In our previous work \cite{LeiTIT}, we demonstrated the capacity optimality of AMP using the I-MMSE and area properties \cite{Guo2005, Bhattad2007}. As mentioned earlier, AMP is particularly suitable to systems with IIDG sensing matrices. However, the OAMP discussed in this paper is applicable to a wider range of applications than AMP since it can handle unitarily invariant matrices, which include IIDG matrices as a special case. This is primarily due to the complicated transfer functions of OAMP, which introduce challenges in analyzing the area property and achieving rate (or capacity optimality). The introduction of matrix inversion in LE and orthogonalization in both LE and NLE further complicates the analysis compared to AMP. Additionally, orthogonal local estimators are generally not locally MMSE-optimal, rendering the direct application of the I-MMSE property in OAMP infeasible. To tackle these difficulties, we remodel the structure of the OAMP receiver by incorporating a ``double-orthogonal" LE and a local MMSE decoder (NLE). However, this ``double-orthogonal" LE introduces a memory term as well as a complex DISO transfer function. Consequently, the I-MMSE property and the results in \cite{LeiTIT} that are applicable only to SISO transfer curves cannot directly apply to OAMP. To address this new challenge, we develop a SISO VTF that allows us to establish the area properties of LUIS while analyzing the achievable rate of OAMP. For more details, please refer to Section \ref{Sec:rate_OAMP}. To the best of our knowledge, this is the first study to present a low-complexity encoder and decoder that achieve the capacity of LUIS, given the signal distribution $P_X$.

\section{Numeric Results}\label{Sec:SIM}
In this section, we present simulation results of the OAMP receiver in LUIS with QPSK modulation and LDPC coding. The irregular LDPC codes are optimized for curve matching. We will not delve into the detailed code design process, as it closely mirrors Section IV-A of \cite{LeiTIT}. We leave the code design for more complicated high-order modulations as our future work. The simulations involve only one sum-product iteration per OAMP iteration, implying that there is no inner iteration in the sum-product decoding for LDPC codes. This simplifies the complexity of the OAMP receiver, as inner iterations could introduce additional computational complexity.

 \emph{Generation of Ill-Conditioned Matrix:}
The matrix $\bm{A}$ is generated by $\bm{A}=\bm{U} \bm{\Sigma V}$.  The eigenvalues $\{d_i\}$ in $\bf{\Sigma}$ are determined by: $d_i/d_{i+1}=\kappa^{1/
T}$ for $i = 1,\ldots, T-1$ and $\sum_{i=1}^Td_i^2=N$, where $T=\min\{M, N\}$ \cite{Vila2014}. Here, $\kappa\ge1$ controls the condition number of $\bm{A}$. The matrices $\bm{U}$ and $\bm{ V}$ are generated using the orthogonal matrices obtained from the QR decomposition of two IIDG matrices. 

\begin{figure*}[t]  
  \centering
  \includegraphics[width=15cm, height=4.3cm]{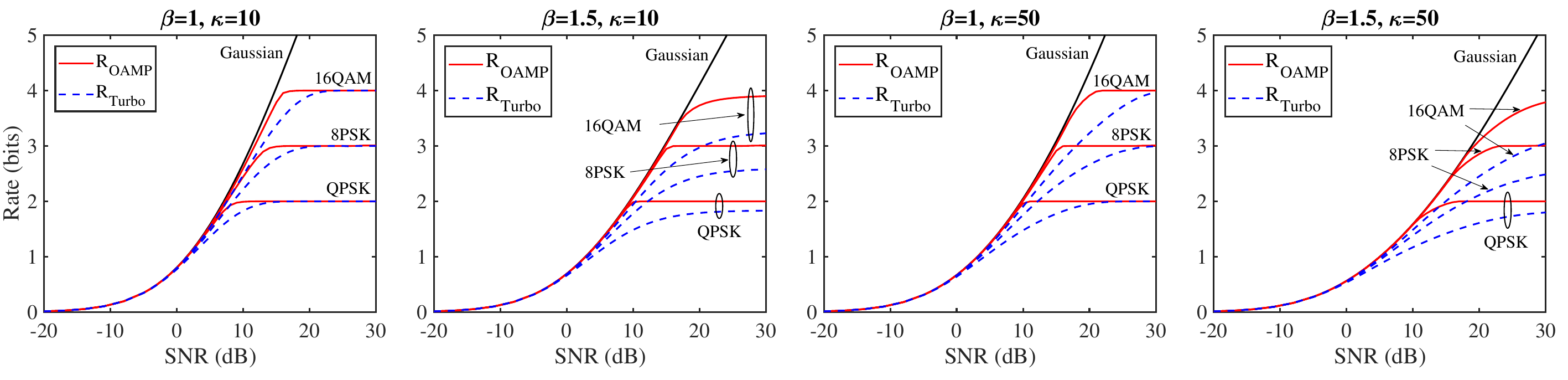}\\  
  \caption{Achievable rate comparison between Turbo and OAMP in LUIS with $\beta=N/M=\{1, 1.5\}$ and $\kappa=\{10, 50\}$ (condition numbers). QPSK, 8PSK, 16QAM and Gaussian modulations are considered. Matrix $\bf{A}$ is generated in the same way as that in Section \ref{Sec:SIM}.}\label{Fig:Rate_OAMP_Turbo}   
\end{figure*} 

\subsection{Comparison with Cascading OAMP Receiver}
Fig. \ref{Fig:Rate_CDD} compares the achievable rates of CAS-OAMP \cite{Guo2005random, Tanaka2002} and OAMP. As can be seen,  OAMP outperforms CAS-OAMP, and their gap increases with $\beta$ and $\kappa$ (condition number). It should be noted that CAS-OAMP exhibits rate jumps due to the discontinuous nature of the first fixed point of OAMP. Please refer to \cite{LeiTIT} for more details.

\subsection{Optimization of Irregular LDPC Codes for OAMP}  
Fig. \ref{Fig:BER_OAMP} presents the BER simulations for the LUIS with optimized irregular LDPC codes \cite{Yuan2008Low, Chung2001}. The ``OAMP-Irreg'' receiver is shown in Fig. \ref{Fig:OAMP}, where $\hat{\phi}$ is a standard sum-product APP decoder. The system sizes are $(N, M)=(500, 500) \;\mr{and}\; (500, 333)$ with channel loads $\beta=\{1, 1.5\}$, respectively. The ill-conditional numbers are $\kappa=\{10, 50\}$.  See Table~\ref{Opt_degree1} for the detailed code parameters. The thresholds of the optimized irregular LDPC codes are quite close to the limits, with a gap of 0.1 dB to 0.2 dB. We consider the optimized irregular LDPC codes for QPSK modulation. Code rate $\approx0.5$, and symbol rate $\approx1$ bits,  resulting in a sum rate $\approx N$ bits per channel use. The number of iterations is less than $250$. As illustrated in Fig.~\ref{Fig:BER_OAMP}, for different $\beta$ and $\kappa$, the BER curves of the optimized irregular LDPC codes at $10^{-5}$ are about $1$~dB away from their respective limits.

\emph{Comparison with Un-Optimized Regular LDPC Code \cite{MaTWC}:} 
We compare the proposed OAMP method, which utilizes optimized irregular codes, with the conventional OAMP approach, which employs unoptimized regular (3, 6) LDPC codes, denoted as ``OAMP-(3, 6)" \cite{MaTWC, Gallager1962}. ``OAMP-(3, 6)'' coincides with $R_{\mr{CAS}}$ in Section \ref{Sec:Comparisons}. As depicted in Fig.~\ref{Fig:BER_OAMP}, when targeting a BER of $10^{-5}$, the proposed OAMP with optimized irregular LDPC codes demonstrates superior performance, exhibiting gains ranging from 0.8 dB to 4 dB compared to ``OAMP-(3, 6)'' for $\beta=\{1, 1.5\}$ and $\kappa=\{10, 50\}$. These results highlight the significant performance improvement achieved through code optimization for OAMP.

\newcommand{\tabincell}[2]{\begin{tabular}{@{}#1@{}}#2\end{tabular}}\renewcommand\arraystretch{1.15}
\begin{table*}[t]  
\caption{Parameters of the optimized Irregular LDPC Codes for  OAMP and Turbo}\label{Opt_degree1}\vspace{-0.3cm}\scriptsize
\centering\setlength{\tabcolsep}{1mm}{
\begin{tabular}{|c||c|c|c|c|c|c|c|c|}
\hline
Methods&   \multicolumn{4}{c|}{ OAMP} & \multicolumn{2}{c|}{Turbo}\\
\hline
$\it{\kappa}$ & \multicolumn{2}{c|}{10}  &  \multicolumn{2}{c|}{50} & 10& 50\\
\hline
$\it{\beta}$&  {$1$} & {$1.5$}& {$1$} &\multicolumn{3}{c|}{1.5}\\
\hline
$\textit{$N$}$ &   \multicolumn{6}{c|}{500}\\
\hline
$\textit{$M$}$  & {$500$} & {$333$}& {$500$} & \multicolumn{3}{c|}{333}\\
\hline
{\tabincell{c}{Codeword\vspace{-0.05cm}\\ length}}  & \multicolumn{6}{c|}{${10^5}^{\;}$}\\
\hline
 {\tabincell{c}{Target\vspace{-0.05cm}\\ code rate}} &\multicolumn{6}{c|}{$0.5$}\\
\hline
{\tabincell{c}{Designed\vspace{-0.05cm}\\ code rate}}  & 0.5087 & 0.5062 &0.5075 &0.4721 &0.5008 &0.48635\\
\hline
${\textit{$R_{\cal C}$}}$   & 1.0178 & 1.0124 &1.0150 &0.9442 &1.0016&0.9727\\
\hline
${\textit{R}}_{\text{sum}}$  & 508.9 & 506.2 &507.5 &472.1 &500.8&486.4\\
\hline
Iterations & \multicolumn{6}{c|}{$250$}\\
\hline
{\tabincell{c}{Check edge\vspace{-0.05cm}\\ distribution}} &\multicolumn{1}{c|}{${\it{\eta}}_{\text{7}}=1$} &${\it{\eta}}_{\text{8}}=1$ &${\it{\eta}}_{\text{8}}=1$ &${\it{\eta}}_{\text{7}}=1$&${\it{\eta}}_{\text{8}}=1$&{\tabincell{c}{ ${\it{\eta}}_{\text{6}}=0.2$ \vspace{-0.05cm}\\ ${\it{\eta}}_{\text{12}}=0.8$}}\\
\hline
  & $\lambda_{2}\!=\!0.3707$    &$\lambda_2\!=\!0.4233$     &$\lambda_{2}\!=\! 0.4624$ &$\lambda_{2}\!=\! 0.5082 $&$\lambda_{2}\!=\!0.4400$    &$\lambda_2\!=\!0.3377$ \\
  & $\lambda_{3}\!=\!0.2329$    &$\lambda_3\!=\!0.0677$     &$\lambda_{3}\!=\! 0.0028$ &$\lambda_{21}\!=\! 0.3238 $&$\lambda_{10}\!=\!0.0577$   &$\lambda_{12}\!=\!0.0481$ \\ Variable
 & $\lambda_{9}\!=\!0.1815$    &$\lambda_{15}\!=\!0.0053$  &$\lambda_{14}\!=\! 0.1924$ &$\lambda_{22}\!=\! 0.0002 $&$\lambda_{11}\!=\!0.2256$   &$\lambda_{13}\!=\!0.2288$\\ edge
  & $\lambda_{10}\!=\!0.0002$   &$\lambda_{16}\!=\!0.2586$  &$\lambda_{15}\!=\! 0.0743$ &$\lambda_{130}\!=\! 0.0001 $& $\lambda_{50}\!=\!0.0401$   &$\lambda_{60}\!=\!0.1012$ \\ distribution
  & $\lambda_{27}\!=\!0.0003$   &$\lambda_{80}\!=\!0.1426$  &$\lambda_{70}\!=\! 0.1649$ &$\lambda_{140}\!=\! 0.0005 $&$\lambda_{60}\!=\!0.1665$  &$\lambda_{70}\!=\!0.1454$\\
  & $\lambda_{28}\!=\!0.1516$   &$\lambda_{200}\!=\!0.1025$ &$\lambda_{150}\!=\! 0.0322$ &$\lambda_{150}\!=\! 0.0018 $ &$\lambda_{250}\!=\!0.0298$  &$\lambda_{300}\!=\!0.1388$\\
 & $\lambda_{29}\!=\!0.0620$   & &$\lambda_{200}\!=\! 0.0711$    &$\lambda_{200}\!=\! 0.1652$&$\lambda_{300}\!=\!0.0403$ &\\
 & $\lambda_{30}\!=\!0.0005$   & & & &&\\
\hline
$(\mathrm{SNR})^{\it{\ast}}_{\text{dB}}$  & $1.7$ & $2.87$ & $3.2$ & 5.2 &$3.8$ &  $7.2$\\
\hline
$\text{Replica capacity}$   &  $1.55$ &  $2.85$  &  $3.15$  & $5.03$&$3.78$ &  $7.12$\\
\hline
\end{tabular}}
\end{table*}

\subsection{Comparison Between OAMP and Turbo} 
Fig. \ref{Fig:Rate_OAMP_Turbo} compares the achievable rates of the conventional Turbo \cite{Yuan2014} and OAMP. Both Turbo and OAMP achieve the Gaussian capacity under Gaussian signaling. However, for QPSK, 8PSK, and 16QAM modulations, OAMP demonstrates replica capacity optimality when it has a unique fixed point, while Turbo exhibits a rate loss and is therefore considered to be capacity sub-optimal. Consequently, OAMP outperforms Turbo, aligning with the findings in \cite{MaTWC}. Furthermore, the rate loss of Turbo increases with higher values of $\beta$ and $\kappa$, and remains negligible if $\beta$ and $\kappa$ are relatively small.  

\begin{figure}[t]  
  \centering
  \includegraphics[width=8.5cm]{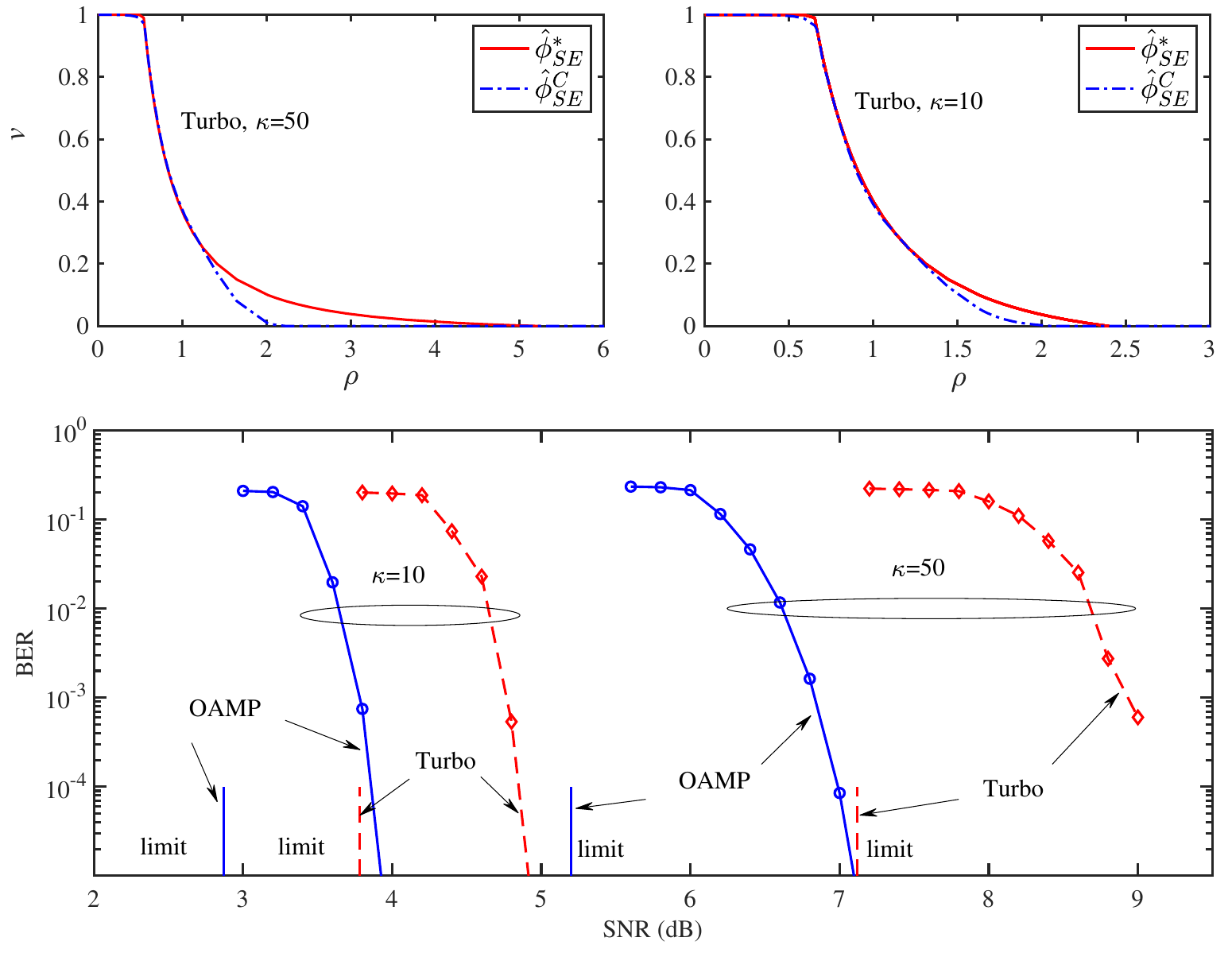}\\
  \caption{Curve matching of Turbo, and BER comparison of OAMP and Turbo   \cite{YC2018TWC, Lei20161b} with optimized irregular LDPC codes, where ``$\hat{\phi}_{\rm{SE}}^*$'' is the target curve, and ``$\hat{\phi}_{\rm{SE}}^{\mathcal{C}}$'' the optimized curve,  ``limit" the rate limits of OAMP or Turbo. QPSK modulation,  codeword length = $10^5$, code rate $\approx$ 0.5,  iterations = $250$, and $N=500$ and $M=333$. See Table \ref{Opt_degree1} for the details.}\label{Fig:BER_OAMP_Turbo}  
\end{figure}

The upper two sub-figures in Fig.~\ref{Fig:BER_OAMP_Turbo} illustrate the transfer curve matching of Turbo \cite{ YC2018TWC, Lei20161b}, while the curve matching of OAMP can be found in Fig. \ref{Fig:BER_OAMP}. Note that the VTF-NLE of OAMP outputs  \emph{a-posteriori} variance, whereas the NLE of Turbo outputs  \emph{extrinsic}  variance. We consider a QPSK-modulated LUIS with system sizes of $(N, M)=(500, 333)$ and condition numbers $\kappa={10, 50}$. The limits of OAMP and Turbo for rate $R=1$ are respectively $2.85$ dB and $3.78$ dB for $\kappa=10$, and $5.03$ dB and $7.12$ dB for $\kappa=50$.  The third sub-figure in Fig.~\ref{Fig:BER_OAMP_Turbo} compares the BERs of the optimized OAMP and the optimized Turbo \cite{ YC2018TWC, Lei20161b} (with 250 iterations). The detailed parameters can be found in Table~\ref{Opt_degree1}, which illustrates that the decoding thresholds are within a range of 0.1 dB to 0.2 dB from their respective limits. In addition, the simulated BERs of OAMP and Turbo are about 1dB for $\kappa=10$ and 2 dB for $\kappa=50$ away from their limits respectively. Compared with the Turbo, OAMP exhibits a 1 dB improvement for $\kappa=10$ and 2 dB improvement for $\kappa=50$ in BER. Overall, the conventional Turbo experiences a significant performance loss in general discrete linear systems, particularly in scenarios involving high transmission rates and/or high condition numbers. On the other hand, OAMP has the capability to approach the replica constrained capacity of discrete systems through proper code design (see Fig. \ref{Fig:BER_OAMP} also for more simulation results).

\emph{Complexity:}
 OAMP and Turbo are comparable in complexity, i.e., $\mathcal{O}\left((\Xi_{LD}\!+\\ \Xi_{NLE})N_{ite}\right)$, where $N_{ite}$ is the number of iterations,  $\Xi_{LD}$ the complexity of LE, and $\Xi_{NLE}$ the complexity of NLE. For LDPC decoding, $\Xi_{NLE}\approx 4\bar{d}_vN_{\cal C}$, where $N_{\cal C}$ is the codeword length, $\{\lambda_i\}$  the degree distribution of variable node, and $\bar{d}_v=\big(\sum_{i}\lambda_i/i\big)^{-1}$  the averaged degree of variable node. On the other hand, the complexity of LE is  $\Xi_{LD}=\mathcal{O}(MN^2)$ due to the matrix inverse operation.  
 
 \section{Conclusion}
An OAMP receiver is considered for a coded LUIS with a unitarily invariant sensing matrix and an arbitrary input distribution. A universal GSO is discussed for OAMP. Our analysis shows that achieves replica capacity optimality when a matched Lipschitz-continuous decoder is used and the state evolution of OAMP converges to a unique fixed point. Specifically, LUIS with Gaussian signaling is studied as a special case, where the replica method is rigorous and the unique fixed-point condition strictly holds. Several area properties are established based on the VTF of OAMP. Furthermore, a curve-matching coding principle is developed for OAMP by utilizing the proposed SISO VTF. Simulation results are provided to verify that the OAMP with optimized irregular LDPC codes approaches the {replica} constrained capacity of LUIS, and significantly outperforms ($0.75$ dB $\sim$ $4$ dB gain) the un-optimized case. Furthermore, we find that the OAMP exhibits a significant improvement in BER performance compared to the state-of-the-art Turbo-LMMSE.

\appendices 

\section{MMSE and Constrained Capacity Conjectured by Replica Method}\label{APP:replica}

\emph{1) MMSE Conjectured by Replica Method}:  Define the \emph{R-transform} \cite{Tulino2004} of a positive semidefinite matrix $\bf{R}$ as 
\BS\label{Eqn:R_transform}\BE
    \mathcal{R}_{\bf{R}}(w) =  \mathcal{S}^{-1}_{\bf{R}}(-w)-1/z,  
\EE
where $\mathcal{S}^{-1}_{\bf{R}}(\cdot)$ is the inverse of the Stieltjes transform  \cite{Tulino2004}:
\BE
    \mathcal{S}_{\bf{R}}(w) = \tfrac{1}{N}{\rm tr}\{(\bf{R}-w\bf{I})^{-1}\}. 
\EE\ES
Let ${\rm mmse}^{-1}_{\cal S}(v)$ be the inverse of $v={\rm mmse}\{\bf{x}|\sqrt{\rho}\bf{x}+\bf{z},\Phi_{\cal S}\}$, where $\bf{z}\sim\mathcal{CN}(\bf{0},\bf{I})$ is independent of $\bf{x}$, and $\rho$ denotes the effective signal-to-noise ratio. The following is an argument made in \cite{Tulino2013, Kabashima2006} using replica symmetric analysis that the MMSE $v^*$ of LUIS satisfies the fixed point equation:  
\BE\label{Eqn:MMSE_replica}
   {\rm mmse}^{-1}_{\cal S} (v^*) = \sigma^{-2}  \mathcal{R}_{\bf{A}^{\rm H}\bf{A}}\left(-\sigma^{-2} v^* \right). 
\EE

\emph{2) Constrained Capacity Conjectured by Replica Method}: Lemma \ref{Pro:dis_cap} below gives the constrained capacity of LUIS predicted by the replica method. 

\begin{lemma} [Replica Constrained Capacity] \label{Pro:dis_cap}
 Suppose that $\rho^*$ is the unique solution of $\rho^* = {\rm SNR}\,\mathcal{R}_{\bf{R}}\big(-{\rm SNR}\,v^*\big)$ where $v^*={\rm mmse}\{\bf{x}|\sqrt{\rho^*}\bf{x}+\bf{z},\Phi\}$ denotes the replica MMSE of a LUIS \cite{Tulino2013, Kabashima2006, Barbier2018b}. The replica constrained capacity\footnote{The replica constrained capacity in \eqref{Eqn:dis_cap_ref} is equivalent to that in \cite{Barbier2018b}.  First, $\rho$, ${\rm SNR}$ and $v^*$ in \eqref{Eqn:dis_cap_ref} correspond to $r$, $\lambda$ and $\mr{E}$ in \cite{Barbier2018b} respectively. Second, the factor $1/2$ for a real LUIS in \cite{Barbier2018b} is removed in \eqref{Eqn:dis_cap_ref} for a complex LUIS.} of a LUIS  \cite{Tulino2013, Kabashima2006, Barbier2018b} is given by  
\BE\label{Eqn:dis_cap_ref}
    C_{\rm Rep}({\rm SNR}) = \int_0^{v^*{\rm SNR}}\!\!\!\!\!\!\!\!\!\!\!\! \mathcal{R}_{\bf{R}}(-z)dz + I(x;\sqrt{\rho^*}x+z)  - \rho^*v^*, 
\EE
where  $\mathcal{R}_{\bf{R}}(\cdot)$ is the \emph{R-transform} of  $\bf{R}=\bf{A}^H\bf{A}$. 
\end{lemma}    

\emph{Note}: The replica method is heuristic as it relies on an unjustified exchange of limits and an unproven replica symmetry assumption. For general right-unitarily invariant $\bf{A}$, the exact constrained capacity of LUIS still remains an open issue. Nonetheless, recent progress has been made in understanding the applicability of the replica method. It was proved that the replica method is correct for IIDG matrices \cite{Reeves_TIT2019, Barbier2017arxiv}, as well as a specific sub-class of LUIS with matrices $\bf{A} = \bf{U} \bf{W}$ \cite{Barbier2018b}, where $\bf{W}$ are IIDG, $\bf{U}$ is a product of a finite number of independent matrices, each with IID matrix-elements that are either bounded or standard Gaussian, and $\bf{U}$ and $\bf{W}$ are independent. Furthermore, recent research \cite{Fan2022} demonstrated the correctness of the replica method for arbitrary rotationally-invariant designs $\bf{A}$, subject to a ``high-temperature" condition that restricts the range of eigenvalues of $\bf{A}^{\rm H}\bf{A}$.

\section{Proof of \texorpdfstring{\eqref{Eqn:MMSE_df}}{TEXT}}\label{APP:Pro_EP}
Let $\bf{\hat{f}}\equiv\hat{f}({\bf{x}}_{\mr{in}}) $. From the orthogonal principle \cite{Williams2001} of MMSE estimation,   
\BS\label{Eqn:MMSE_orth}\begin{align}
&{\rm{E}}\big\{ {\bf{x}}_{\mr{in}}^{\mr{H}} (\bf{\hat{f}} - \bf{x})  \big\}  = 0,\label{Eqn:MMSE_ortha}\\
&{\rm{E}}\big\{  \bf{\hat{f}}^{\mr{H}} (\bf{\hat{f}}  - \bf{x}) \big\} = 0.  \label{Eqn:MMSE_orthb}
\end{align}\ES
Assume $\bf{\hat{f}}=\alpha_{\hat{f}}\bf{x}+\bf{\xi}_{\hat{f}}$ with ${\rm{E}}\{ \bf{x}^{\mr{H}} \bf{\xi}_{\hat{f}}\}= 0$. We further have   
\BS\label{Eqn:MMSE_vars}\begin{align}
  {\mr{E}}\big\{ \bf{\xi}_{\mr{in}} ^{\mr{H}}  \bf{\hat{f}} \big\} &={\rm{E}}\big\{\bf{\xi}_{\mr{in}}^{\mr{H}}\bf{\xi}_{\hat{f}} \big\}   =   (1 - \alpha _{\hat{f}})\mr{E}\big\{\|{\bf{x}}\|^2\big\},\label{Eqn:MMSE_varsb}\\
 \!\!\!{\rm{E}}\big\{ { \| {\bf{\hat{f}} \! - \!\bf{x}}  \|^2}\big\} &\!=\! {\rm{E}}\big\{ \!- {\bf{x}^{\mr{H}}}( \bf{\hat{f}} \! -\! \bf{x})\big\}\! =\! (1 \!-\! \alpha _{\hat{f}})\mr{E}\big\{\|{\bf{x}}\|^2\big\},\label{Eqn:MMSE_varsc}
\end{align}\ES
where \eqref{Eqn:MMSE_varsb} follows \eqref{Eqn:MMSE_ortha} and $\mr{E}\big\{\bf{\xi}_{\rm in}^{\rm H}{\bf{x}}\big\}=0$, and \eqref{Eqn:MMSE_varsc} follows \eqref{Eqn:MMSE_orthb} and $\mr{E}\big\{\bf{\xi}_{\hat{f}}^{\rm H}{\bf{x}}\big\}=0$. From \eqref{Eqn:MMSE_vars}, we have 
\BE\label{Eqn:MMSE_Bup}
\tfrac{1}{N}{\mr{E}}\big\{ \bf{\xi}_{\mr{in}} ^{\mr{H}}  \bf{\hat{f}} \big\} = \tfrac{1}{N}{\rm{E}}\big\{   \| \bf{\hat{f}} \!-\! \bf{x}   \|^2\big\}=v_{\hat{f}}. 
\EE
From \eqref{Eqn:MMSE_Bup} and \eqref{Eqn:GSO_B}, we obtain \eqref{Eqn:MMSE_df}. 


\section{Proof of Theorem \ref{The:area_LUIS}}\label{APP:area_express} 
To establish Theorem \ref{The:area_LUIS}, our proof proceeds in two main steps. Firstly, we establish the area expression of $A_{\rm ADGO}$ as given in \eqref{Eqn:A_C}. Subsequently, we demonstrate the area property $A_{\rm ADGO} = C_{\rm Rep}({\rm SNR})$. Prior to proving $A_{\rm ADGO} = C_{\rm Rep}({\rm SNR})$, we establish the consistency of the fixed points in \eqref{Eqn:A_C} and \eqref{Eqn:dis_cap_ref}. Additionally, we validate the consistency between $C_{\rm Rep}({\rm SNR})$ and $A_{\rm ADGO}$.

\subsection{Proof of the Area Expression in \texorpdfstring{\eqref{Eqn:A_C}}{TEXT}}
Eqn. \eqref{Eqn:A_C1} is straightforward. We next prove \eqref{Eqn:A_C2}. Let $\bf{D}(s) = s\bf{I} + {\rm SNR}\bf{A}^H\bf{A}$. Then
\BS\label{e42}\begin{align} 
 & \int_{\rho^*}^{{\rm SNR}}\eta^{-1}_{\rm SE}(\rho) d\rho \\
&= \int_{v=v^*}^{v=0}   v\,  d \left( v^{-1} -  \big[\hat{\gamma}^{- 1}_{\rm SE}(v)\big]^{- 1}  \right)    \label{Eqn:AC_a}  \\
&= \big[ -\log v \big]_{v=v^*}^{v=0}  - \int_{v=v^*}^{v=0}  v \, d\, \big[\hat{\gamma}^{- 1}_{\rm SE}(v)\big]^{- 1}   \label{Eqn:AC_b}  \\
&= \big[ -\log v \big]_{v=v^*}^{v=0}  - \int_{[v^*]^{-1}-\rho^*}^{\infty}  \hat{\gamma}_{\rm SE}(s^{- 1}) \, d\,s  \label{Eqn:AC_c}  \\
&=    \big[ -\log v \big]_{v=v^*}^{v=0} -   \int_{[v^*]^{-1}-\rho^*}^{\infty}  {  \tfrac{1}{N}\mathrm{tr}\big\{ [\bf{D}(s)]^{\!-1}\big\} d s}  \label{Eqn:AC_d} \\
&= \big[ -\log v \big]_{v=v^*}^{v=0} - \left[{ \tfrac{1}{N} \log \det\big( \bf{D}(s) \big)}\right]^{s=\infty}_{s=[v^*]^{-1}-\rho^*}  \label{Eqn:AC_e} \\ 
&= \log  v^* +  \!{ \tfrac{1}{N} \log \det\big( \bf{B}(\rho^*, v^*) \big)},\label{Eqn:AC_f}
\end{align} \ES
where  \eqref{Eqn:AC_a} follows $\eta_{\rm SE}(v)=v^{-1} -  \big[\hat{\gamma}^{- 1}_{\rm SE}(v)\big]^{- 1}$, \eqref{Eqn:AC_c} follows $s=\big[\hat{\gamma}^{- 1}_{\rm SE}(v)\big]^{- 1}$, $\hat{\gamma}^{- 1}_{\rm SE}(0)=0$ and the fix-point equation $\big[\hat{\gamma}^{- 1}_{\rm SE}(v^*)\big]^{- 1} = [v^*]^{-1}-\rho^*$, \eqref{Eqn:AC_d} follows $\int {\mathrm{tr}\{ (s\bf{I} + \bf{A})^{-1}\} ds }= \log\det (s\bf{I} + \bf{A})$, and \eqref{Eqn:AC_f} follows $-\lim\limits_{v\to0}\log v -\lim\limits_{s\to\infty}{ \tfrac{1}{N} \log \det\big( \bf{D}(s) \big)}  =0$.  Therefore, we obtain the desired \eqref{Eqn:A_C2} from \eqref{Eqn:A_C1} and \eqref{e42}. 

\subsection{Proof of the Area Property \texorpdfstring{$A_{\rm ADGO} = C_{\rm Rep}({\rm SNR})$}{TEXT}}\label{APP:cap_opt} 
We now show $A_{\rm ADGO} = C_{\rm Rep}({\rm SNR})$, where $A_{\rm ADGO}$ is given in \eqref{Eqn:A_C} and $C_{\rm Rep}({\rm SNR})$ is given in \eqref{Eqn:dis_cap_ref}.  First, we demonstrate the consistency of the fixed points\footnote{The consistency of the fixed points of OAMP/VAMP with the replica MMSE was also proved in \cite{Ma2016, Rangan2016}.} in \eqref{Eqn:A_C} and \eqref{Eqn:dis_cap_ref}. Then, we demonstrate the consistency between $C_{\rm Rep}({\rm SNR})$ and $A_{\rm ADGO}$.

\subsubsection{Consistency of the Fixed Points} The following proof is based on an identity 
\BE\label{Eqn:RS_T}
\mathcal{R}_{\bf{R}}(z)=\mathcal{S}^{-1}_{\bf{R}}(-z)-{z}^{-1}, 
\EE
where $\mathcal{S}^{-1}_{\bf{R}}(\cdot)$ is the inverse of the \emph{Stieltjes transform } $\mathcal{S}_{\bf{R}}(z)=\mr{E}_{\lambda_{\bf{R}}}\big\{1/(\lambda_{\bf{R}}-z)\big\}$. Recall that $\rho^*$ in \eqref{Eqn:A_C} is the solution of 
\BS\label{Eqn:our_rho}\BE 
    \eta^{-1}_{\rm SE}(\rho)=\hat{\phi}^{\cal S}_{\rm SE}(\rho), 
\EE
where $\eta^{-1}_{\rm SE}(\cdot)$ is the inverse of  $\eta_{\rm SE}(v) \equiv v^{-1}-[\hat{\gamma}_{\rm SE}^{-1}(v)]^{-1}$ and $\hat{\gamma}_{\rm SE}^{-1}(\cdot)$ is the inverse of   
\begin{align}
  \hat{\gamma}_{\rm SE}(v) &= \tfrac{1}{N}{\mr{tr}}\big\{ [{\rm SNR}\bf{A}^H\bf{A} + v^{-1}\,\bf{I}]^{-1}\big\}\nonumber\\
&={\rm SNR}^{-1}\mathcal{S}_{\bf{R}}\big(-({\rm SNR}\,v)^{-1}\big).  \label{Eqn:gamma_mmse2}
\end{align} \ES
Eqn. \eqref{Eqn:our_rho} can be rewritten to  
\BE\label{Eqn:our_rho2}
\hat{\phi}_{\rm SE}^{\cal S}(\rho)=  \hat{\gamma}_{\rm SE} \Big( \big( [\hat{\phi}_{\rm SE}^{\cal S}(\rho)]^{-1}-\rho \big)^{-1}\Big). 
\EE
Substituting \eqref{Eqn:gamma_mmse2} into \eqref{Eqn:our_rho2}, we have  
\BE\label{Eqn:inter_eqn}
    {\rm SNR} \, \hat{\phi}_{\rm SE}^{\cal S}(\rho) = \mathcal{S}_{\bf{R}}\Big(-{\rm SNR}^{-1}\big( [\hat{\phi}^{\cal S}_{\rm SE}(\rho)]^{-1}-\rho \big) \Big). 
\EE 
Taking the inverse $\mathcal{S}^{-1}_{\bf{R}}$ at both sides of \eqref{Eqn:inter_eqn}, we have 
\BE
 \mathcal{S}^{-1}_{\bf{R}}\big({\rm SNR} \, \hat{\phi}_{\rm SE}^{\cal S}(\rho)\big) = -{\rm SNR}^{-1}\big( [\hat{\phi}^{\cal S}_{\rm SE}(\rho)]^{-1}-\rho \big). 
\EE
Using \eqref{Eqn:RS_T}, we obtain  
\BE
\rho={\rm SNR}\,\mathcal{R}_{\bf{R}}\big(-{\rm SNR}\, \hat{\phi}_{\rm SE}^{\cal S}(\rho)\big), 
\EE
which is the same as the fixed point equation in \eqref{Eqn:A_C}, i.e., $\rho^*$ in \eqref{Eqn:dis_cap_ref} and \eqref{Eqn:A_C} are the same.

\subsubsection{Consistency of \texorpdfstring{$ C_{\rm Rep}({\rm SNR})$ and $A_{\rm ADGO}$}{TEXT}}
Recall \eqref{Eqn:A_C}: 
\BE
 A_{\rm ADGO} =\int_{0}^{\rho^*}  {\hat{\phi}_{\rm SE}^{\cal S}(\rho) \,d\rho} + \int_{\rho^*}^{{\rm SNR}}\eta^{-1}_{\rm SE}(\rho) d\rho. 
\EE
Based on the \emph{I-MMSE Lemma}, we have 
\BE\label{Eqn:part1}
 \int_{0}^{\rho^*} {\hat{\phi}_{\rm SE}^{\cal S}(\rho) d\rho} = I(x;\sqrt{\rho^*}x+z). 
\EE
Furthermore, 
\BS\label{Eqn:part2}\begin{align}
 &  \int_{\rho^*}^{{\rm SNR}}\eta^{-1}_{\rm SE}(\rho) d\rho \\
&= \int_{v=v^*}^{v=0}   v\,  d \left( v^{-1} -  \big[\hat{\gamma}^{- 1}_{\rm SE}(v)\big]^{- 1}  \right)  \label{Eqn:part2a} \\
&= \int_{s=[v^*]^{-1}-\rho^*}^{s=\infty}   \hat{\gamma}_{\rm SE}(s^{-1})\,  d \left( [\hat{\gamma}_{\rm SE}(s^{-1})]^{-1} -  s  \right)    \label{Eqn:part2b}  \\%
 &=\int_{s=[v^*]^{-1}-\rho^*}^{s=\infty}\!\!\!\!\!\!\!\!\!\!\!\!\!\!\!\!\!\!\!\!\!\!\!\! \mathcal{S}_{\bf{R}}\big(\!\!-\!{\rm SNR}^{-1} s\big) d   \left( \! \big[\mathcal{S}_{\bf{R}}\big(\!\!-\!{\rm SNR}^{-1} s\big)\big]^{-1}\!\! -\!{\rm SNR}^{-1}  s \!\right) \label{Eqn:part2c}\\%
 &=-\int_{t=-\infty}^{t=-{\rm SNR}^{-1}([v^*]^{-1}-\rho^*)}\mathcal{S}_{\bf{R}}(t)d    \left(  \mathcal{S}_{\bf{R}}(t)^{-1} +t  \right)\label{Eqn:part2d}\\
 &=-\int_{0}^{{\rm SNR}\,v^*}  z\, d  \mathcal{R}_{\bf{R}}(-z) \label{Eqn:part2e} \\
 &=\int_{0}^{{\rm SNR}\,v^*}  \mathcal{R}_{\bf{R}}(-z)  d z  +  \left[z\mathcal{R}_{\bf{R}}(-z) \right]_{z={\rm SNR}\,v^*}^{z=0}\label{Eqn:part2f}\\
 &=\int_{0}^{{\rm SNR}\,v^*}  \mathcal{R}_{\bf{R}}(-z)  d z  -  \rho^*v^*,\label{Eqn:part2g}
\end{align}\ES
where  \eqref{Eqn:part2a} follows $\eta_{\rm SE}(\rho)=v^{-1} -  \big[\hat{\gamma}^{- 1}_{\rm SE}(v)\big]^{- 1}$, \eqref{Eqn:part2b} follows $s=\big[\hat{\gamma}^{- 1}_{\rm SE}(v)\big]^{- 1}$, $\hat{\gamma}^{- 1}_{\rm SE}(0)=0$ and the fix-point equation $\big[\hat{\gamma}^{- 1}_{\rm SE}(v^*)\big]^{- 1} = [v^*]^{-1}-\rho^*$, \eqref{Eqn:part2c} follows $\hat{\gamma}_{\rm SE}(v) ={\rm SNR}^{-1}\mathcal{S}_{\bf{R}}\big(-({\rm SNR}\,v)^{-1}\big)$, \eqref{Eqn:part2d} follows $t=-{\rm SNR}^{-1} s$, \eqref{Eqn:part2e} follows $z= \mathcal{S}_{\bf{R}}(t)$, $\mathcal{R}_{\bf{R}}(z)=\mathcal{S}^{-1}_{\bf{R}}(-z)-{z}^{-1}$, $0= \mathcal{S}_{\bf{R}}(-\infty)$ and ${\rm SNR}\, v^*  =\mathcal{S}_{\bf{R}}\big({\rm SNR}^{-1}\, [\rho^*-(v^*)^{-1}] \big)$,  and \eqref{Eqn:part2g} follows  $\rho^* = {\rm SNR}\,\mathcal{R}_{\bf{R}}\big(-{\rm SNR}\,
v^*\big)$. 

From \eqref{Eqn:part1} and \eqref{Eqn:part2}, we have $A_{\rm ADGO} = C_{\rm Rep}({\rm SNR})$. 
  
\section{Proof of Lemma \ref{Lem:same_FP}}\label{APP:same_FP}  
  Substituting ${\eta}_{\rm SE}(v)= v^{-1}-[\hat{\gamma}_{\rm SE}^{-1}(v)]^{-1}$ (see \eqref{Eqn:TF}) into the fixed-point equation $\hat{\phi}_{\rm SE}^{\cal S}(\rho)= {\eta}^{-1}_{\rm SE}(\rho)$, we have 
     \BE
         \rho  ={\eta}_{\rm SE}\big(\hat{\phi}_{\rm SE}^{\cal S}(\rho)\big)
          =[\hat{\phi}_{\rm SE}^{\cal S}(\rho)]^{-1} -[\hat{\gamma}_{\rm SE}^{-1}\big(\hat{\phi}_{\rm SE}^{\cal S}(\rho)\big)]^{-1},   
     \EE
     which can be rewritten to 
     \BE \label{Eqn:FPv1}
         \hat{\phi}_{\rm SE}^{\cal S}(\rho) = \hat{\gamma}_{\rm SE} \Big(\big[ [\hat{\phi}_{\rm SE}^{\cal S}(\rho)]^{-1} - \rho\big]^{-1}\Big).   
     \EE 
     Similarly, substituting  ${\phi}_{\rm SE}^{\cal S}(\rho)= \big([\hat{\phi}_{\rm SE}^{\cal S}(\rho)]^{-1}- \rho \big)^{-1}$ (see \eqref{Eqn:TF1}) into the fixed-point equation ${\phi}_{\rm SE}^{\cal S}(\rho)= {\gamma}^{-1}_{\rm SE}(\rho)$, we have 
     \BE
         {\gamma}^{-1}_{\rm SE}(\rho) =   \big([\hat{\phi}_{\rm SE}^{\cal S}(\rho)]^{-1} - \rho \big)^{-1}. 
     \EE     
    Furthermore, following ${\gamma}_{\rm SE}(\vartheta)= [\hat{\gamma}_{\rm SE}(\vartheta)]^{-1} - \vartheta^{-1}$   (see \eqref{Eqn:TF1}), we then have 
    \BS \begin{align}
         \rho &= {\gamma}_{\rm SE}\Big( \big([\hat{\phi}_{\rm SE}^{\cal S}(\rho)]^{-1} - \rho \big)^{-1} \Big) \\
         &= \Big[\hat{\gamma}_{\rm SE}\Big( \big([\hat{\phi}_{\rm SE}^{\cal S}(\rho)]^{-1} - \rho \big)^{-1} \Big)\Big]^{-1} - [\hat{\phi}_{\rm SE}^{\cal S}(\rho)]^{-1} - \rho,
     \end{align}\ES 
    which can be further simplified to 
    \BE\label{Eqn:FPv2}
         \hat{\phi}_{\rm SE}^{\cal S}(\rho) = \hat{\gamma}_{\rm SE} \Big(\big[ [\hat{\phi}_{\rm SE}^{\cal S}(\rho)]^{-1} - \rho\big]^{-1}\Big).   
   \EE 
    which is the same as \eqref{Eqn:FPv1}. 

\section{Proof of Lemma \ref{Lem:ZF_equal}}\label{APP:ZF_equal}
Since ${\gamma}_{\rm SE}(\cdot)$ is a strictly decreasing function \cite[Lemma 2]{Ma2016}, then ${\phi}_{\rm SE}^{\cal C}(\rho)<{\gamma}^{-1}_{\rm SE}(\rho)$ is equivalent to  
   \BE\label{Eqn:SE_ineq}
      {\gamma}_{\rm SE} \big( {\phi}_{\rm SE}^{\cal C}(\rho) \big) > \rho. 
   \EE
   Following ${\phi}_{\rm SE}^{\cal C}(\rho)= \big([\hat{\phi}_{\rm SE}^{\cal C}(\rho)]^{-1}- \rho \big)^{-1}$ and ${\gamma}_{\rm SE}(\vartheta)= [\hat{\gamma}_{\rm SE}(\vartheta)]^{-1} - \vartheta^{-1}$   (see \eqref{Eqn:TF1}), we have  
   \BS\label{Eqn_ieqn}\begin{align}
     & {\gamma}_{\rm SE} \big( {\phi}_{\rm SE}^{\cal C}(\rho) \big) \\
      &= {\gamma}_{\rm SE} \Big(  \big([\hat{\phi}_{\rm SE}^{\cal C}(\rho)]^{-1}- \rho \big)^{-1}  \Big) \\
      &= \Big[\hat{\gamma}_{\rm SE}(\big([\hat{\phi}_{\rm SE}^{\cal C}(\rho)]^{-1}- \rho \big)^{-1})\Big]^{-1} - [\hat{\phi}_{\rm SE}^{\cal C}(\rho)]^{-1} + \rho.
   \end{align}  
   \ES  
   Then, from \eqref{Eqn_ieqn}, we  rewrite \eqref{Eqn:SE_ineq} to  
    \BE\label{Eqn:SE_ineq2}
      \hat{\gamma}_{\rm SE}\Big(\big([\hat{\phi}_{\rm SE}^{\cal C}(\rho)]^{-1}- \rho \big)^{-1}\Big) < \hat{\phi}_{\rm SE}^{\cal C}(\rho), 
   \EE 
   where $\hat{\gamma}_{\rm SE}(\cdot)$ is an MMSE function, which is a strictly  increasing function \cite{Ma2016}. Hence, \eqref{Eqn:SE_ineq2} is equivalent to  
      \BE\label{Eqn:SE_ineq3}
      \big([\hat{\phi}_{\rm SE}^{\cal C}(\rho)]^{-1}- \rho \big)^{-1} < \hat{\gamma}_{\rm SE}^{-1}\big( \hat{\phi}_{\rm SE}^{\cal C}(\rho)\big), 
   \EE  
   which can be rewritten to 
    \BE\label{Eqn:SE_ineq4}
       [\hat{\phi}_{\rm SE}^{\cal C}(\rho)]^{-1} - \big[\hat{\gamma}_{\rm SE}^{-1}\big( \hat{\phi}_{\rm SE}^{\cal C}(\rho)\big)\big]^{-1} >\rho. 
   \EE   
  Since ${\eta}_{\rm SE}(v)= v^{-1}-[\hat{\gamma}_{\rm SE}^{-1}(v)]^{-1}$ (see \eqref{Eqn:TF}), we have  
  \BE\label{Eqn:SE_ineq5}
      [\hat{\phi}_{\rm SE}^{\cal C}(\rho)]^{-1} - \big[\hat{\gamma}_{\rm SE}^{-1}\big( \hat{\phi}_{\rm SE}^{\cal C}(\rho)\big)\big]^{-1} ={\eta}_{\rm SE}( \hat{\phi}_{\rm SE}^{\cal C}(\rho)). 
   \EE   
   Using \eqref{Eqn:SE_ineq5}, we rewrite \eqref{Eqn:SE_ineq4} to 
    \BE\label{Eqn:SE_ineq6}
       {\eta}_{\rm SE}( \hat{\phi}_{\rm SE}^{\cal C}(\rho)) >\rho. 
   \EE   
   From Proposition \ref{Pro:Mon_decr}, ${\eta}_{\rm SE}( \cdot)$ is a strictly  decreasing function. Hence, \eqref{Eqn:SE_ineq6} is equivalent to 
   \BE\label{Eqn:SE_ineq7}
       \hat{\phi}_{\rm SE}^{\cal C}(\rho)< {\eta}_{\rm SE}^{-1}(\rho). 
   \EE 
   Hence, we complete the proof of Lemma \ref{Lem:ZF_equal}.

\end{document}